\documentclass[prd,aps,nofootinbib,tightenlines]{revtex4}
\usepackage{xspace}
\usepackage{mathrsfs}
\usepackage{amsmath}
\usepackage{amssymb}
\usepackage{epsfig}
\usepackage{graphicx}
\usepackage{booktabs}
\usepackage{multirow}
\usepackage{subfigure}
\usepackage{bm}
\usepackage{times}
\usepackage{braket}
\usepackage{color}
\usepackage{slashed}
\usepackage{hyperref}
\usepackage{threeparttable}
\definecolor{Red}{rgb}{1.,0.,0.}
\newcommand{\Red}[1]{{\color{nicered}{#1}}}
\definecolor{Blue}{rgb}{0.,0.,1.}

\definecolor{nicered}{rgb}{0.7,0.1,0.1}
\definecolor{nicegreen}{rgb}{0.1,0.5,0.1}
\hypersetup{colorlinks,citecolor=nicegreen,linkcolor=nicered}
\newcommand{\optbar}[1]{\shortstack{{\tiny (\rule[.4ex]{1em}{.1mm})}\\ [-.7ex] $#1$}}
\newcommand{\beq}{\begin{eqnarray}}
\newcommand{\eeq}{\end{eqnarray}}
\newcommand{\non}{\nonumber\\ }

\newcommand{\KorKbar}{\kern 0.18em\optbar{\kern -0.18em K}{}\xspace}

\begin{document}
\title{Improved global determination of two-meson distribution amplitudes from multi-body $B$ decays
}
\author{Da-Cheng Yan$^1$}
\author{Hsiang-nan Li$^2$}
\author{Zhou Rui$^3$}
\author{Zhen-Jun Xiao$^4$}
\author{Ya Li$^5$}                \email[Corresponding author:]{liyakelly@163.com}

\affiliation{$^1$ School of Mathematics and Physics, Changzhou University, Changzhou, Jiangsu 213164, China}
\affiliation{$^2$ Institute of Physics, Academia Sinica, Taipei, Taiwan 115, Republic of China}
\affiliation{$^3$ Department of Physics, Yantai University, Yantai 264005, China}
\affiliation{$^4$ Department of Physics and Institute of Theoretical Physics, Nanjing Normal University, Nanjing 210023, Jiangsu, China}
\affiliation{$^5$ Department of Physics, College of Sciences, Nanjing Agricultural University,
Nanjing, Jiangsu 210095, China}

\date{\today}

\begin{abstract}
We improve the perturbative QCD (PQCD) formalism for multi-body charmless hadronic $B$ meson decays, such as $B\to VP_3\to P_1P_2P_3$, by resolving the possible discrepancy in parametrizing the contribution of the $P$-wave resonance $V$ to the two-meson distribution amplitudes (DAs) associated with the pairs $P_1P_2=\pi\pi, K\pi, KK$.
The determination of the Gegenbauer moments in the two-meson DAs is then updated in the global fit of the improved PQCD factorization formulas at leading order in the strong coupling $\alpha_s$ to available data for branching ratios and polarization fractions of three- and four-body $B$ decays.
The convergence of the Gegenbauer expansion of the resultant two-meson DAs is manifest.
The satisfactory quality of the fit implies the consistency of the PQCD framework for multi-body $B$ decays and the universality of the nonperturbative two-meson DAs.
In particular, the predicted longitudinal polarization fraction $f_0(B_s^0\to K^{*0} {\bar K}^{*0})=28.2^{+8.8}_{-9.5} \%$ with the updated Gegenbauer moments matches well the measurement.
The observable $L_{K^{*0}{\bar K}^{*0}}=7.7^{+4.9}_{-3.8}$, defined as the ratio of the longitudinal amplitudes of the two $U$-spin related channels $B_s^0\to K^{*0} {\bar K}^{*0}$ and $B^0\to K^{*0} {\bar K}^{*0}$, accommodates the current data within errors.
It is found that the direct $CP$ asymmetries ${\cal A}^{0,||,\bot}_{\rm CP}$ in the polarization states of some four-body decays $B\to V_1V_2\to (P_1P_2)(P_3P_4)$ might be large, but the destruction among them result in small net $CP$ violation.
Our predictions can be confronted with LHCb and Belle-II data in the future.
\end{abstract}

\maketitle


\section{Introduction}

Multi-body $B$ meson decays offer one of the most promising avenues for studying involved strong and weak dynamics and for probing $CP$ violation (CPV).
CPV is a number in two-body decays, but depends on the invariant mass of a meson pair in final states of multi-body decays and varies from region to region in a Dalitz plot~\cite{Dalitz:1953cp,Dalitz:1954cq}.
Sizable direct $CP$ asymmetries in localized regions of the Dalitz plots for the $B^\pm \to \pi^\pm \pi^+ \pi^-$ and $B^\pm \to K^\pm K^+ K^-$ decays have been experimentally well established~\cite{LHCb:2013ptu,LHCb:2013lcl,LHCb:2022fpg}.
The CPV originating from the interference between $S$- and $P$-wave resonances was also observed, which causes large local asymmetries~\cite{LHCb:2019jta}.
Multi-body $B$ meson decays thus deserve thorough theoretical and phenomenological investigations.
Nevertheless, their analysis is considerably more challenging than that of two-body decays owing to complicated interplay between resonant and nonresonant contributions as well as possible significant final-state interactions~\cite{Bediaga:2013ela, Bediaga:2015mia, Kang:2013jaa}.

A factorization formalism that describes multi-body $B$ meson decays in entire phase space is not yet available at present.
It seems reasonable to assume the factorization of three-body $B$ meson decays as a working principle, when two final-state mesons are collimated and the bachelor meson recoils back~\cite{Chen:2002th, El-Bennich:2009gqk, Virto:2016fbw,Krankl:2015fha}.
The nonperturbative (collinear) dynamics responsible for the production of the meson pair, including final-state interactions between the two mesons, is absorbed into two-meson distribution amplitudes (DAs) naturally~\cite{G,G1,DM,Diehl:1998dk,Diehl:1998dk1,Diehl:1998dk2,MP}.
The formulation of three-body $B$ meson decays is then simplified to that of quasi-two-body decays, where a Feynman diagram for hard kernels contains a single virtual gluon exchange at leading order (LO) in the strong coupling $\alpha_s$.
The same idea is applicable to the exploration of four-body charmless hadronic $B$ meson decays, which proceed dominantly with two intermediate resonances.

To attain definite predictions from the above factorization theorem, nonperturbative inputs of two-meson DAs must be known to high precision.
We have performed a global fit of the Gegenbauer moments in two-meson DAs to measured branching ratios and direct $CP$ asymmetries of three-body charmless hadronic $B$ meson decays $B\to VP_3\to P_1P_2P_3$ in the perturbative QCD (PQCD) approach, where $V$ stands for an intermediate vector resonance, and $P_i$, $i=1,2,3$, represent final-state pseudoscalar mesons~\cite{Li:2021cnd}.
The Gegenbauer moments in the DAs for the mesons $P_3=\pi,K$ have been fixed in the global analysis of two-body $B$ meson decays~\cite{Hua:2020usv}. The leading-twist (twist-2) and subleading-twist (twist-3) DAs for the pairs $P_1P_2=\pi\pi, K\pi$ and $KK$ with the intermediate vector mesons $V=\rho, K^*$ and $\phi$, respectively, were then determined in Ref.~\cite{Li:2021cnd}.
It has been demonstrated that the predictions for various observables in multi-body $B$ meson decays from the fitted Gegenbauer moments are in general consistency with data ~\cite{Rui:2021kbn,Li:2021qiw,Zhang:2021nlw,Yan:2022kck,Zhang:2022pfn,Zou:2022xrr,Yan:2023yvx,Yan:2024ymv}.

However, the Gegenbauer moments in the $K\pi$ and $KK$ DAs, being slightly higher than unity~\cite{Li:2021cnd}, are not favored in view of the convergence of the Gegenbauer expansion.
Hence, We introduce additional parameters to compensate the possible discrepancy between two theoretical treatments of the hadronic matrix elements for vacuum transition to meson pairs~\cite{Dedonder:2010fg}; one is based on the definitions for time-like meson form factors, and another is in terms of Breit-Wigner (BW) propagators for intermediate $P$-wave resonances.
With the above new ingredient, we refine the global determination of the Gegenbauer moments, which indeed exhibit better convergence.
We also take this chance to extend the previous analyses solely for three-body $B$ decays~\cite{Li:2021cnd} and for the $\pi\pi$~\cite{Yan:2023yvx} and $KK$~\cite{Yan:2022kck} DAs specifically to a full coverage of the $\pi\pi$, $K\pi$ and $KK$ DAs in three- and four-body $B$ decays.
The branching ratios, $CP$ asymmetries and polarization fractions of numerous three- and four-body $B$ meson decays are then predicted using the updated Gegenbauer moments.
It will be shown that the derived longitudinal polarization fraction $f_0(B_s^0\to K^{*0} {\bar K}^{*0})=28.2^{+8.8}_{-9.5} \%$, distinct from the previous result~\cite{Rui:2021kbn}, agrees well with the measurement now.
The observable $L_{K^{*0}{\bar K}^{*0}}=7.7^{+4.9}_{-3.8}$, defined as the ratio of the longitudinal amplitudes of the two $U$-spin related channels $B_s^0\to K^{*0} {\bar K}^{*0}$ and $B^0\to K^{*0} {\bar K}^{*0}$~\cite{Alguero:2020xca}, accommodates the current data within errors.
It is found that the direct $CP$ asymmetries ${\cal A}^{0,||,\bot}_{\rm CP}$ in the polarization states of the four-body decays $B\to V_1V_2\to (P_1P_2)(P_3P_4)$ might be large, but the destruction among them turn in small net CPV.
Our predictions can be confronted with LHCb and Belle-II data in the future.


The rest of the paper is organized as follows. The kinematic variables for multi-body hadronic $B$ meson decays are assigned in Sec.~II.
The two-meson $P$-wave DAs are parametrized and normalized to time-like form factors, which take the relativistic Breit-Wigner (RBW) model~\cite{LHCb:2018oeg} or the Gounaris-Sakurai (GS) model~\cite{Gounaris:1968mw}.
Such a parametrization is allowed by the Watson theorem \cite{Watson:1952ji}, that guarantees the factorization of elastic rescattering effects into a final-state meson pair.
The extra parameters are associated with the $P$-wave strength in two-meson DAs.
We elaborate how to perform the global fit, and present and discuss the numerical results for branching ratios, polarization fractions and direct $CP$ asymmetries of multi-body $B$ meson decays in Sec.~III, which is followed by the Conclusion.

\section{THEORETICAL FRAMEWORK}\label{sec:2}

\subsection{Kinematics}

Consider the charmless $B$ meson decay into three pseudoscalar mesons via an intermediate vector resonance, $B(p_B)\rightarrow V(p)P_3(p_3)\rightarrow P_1(p_1)P_2(p_2)P_3(p_3)$, with the momenta $p_B=p+p_3$ and $p=p_1+p_2$.
We work in the $B$ meson rest frame and arrange the meson pair $P_1P_2$ and the bachelor meson $P_3$ to move in the directions $n=(1,0,0_{\rm T})$ and $v=(0,1,0_{\rm T})$, respectively.
The above momenta in the light-cone coordinates are chosen as
\begin{eqnarray}
p_{B}&=&\frac{m_{B}}{\sqrt 2}(1,1,\textbf{0}_{\rm T}), ~~~~~~\quad k_{B}=\left(0,x_B \frac{m_{B}}{\sqrt2} ,\textbf{k}_{B \rm T}\right),\nonumber\\
p&=&\frac{m_{B}}{\sqrt2}(f_{+},f_{-},\textbf{0}_{\rm T}), ~\quad k= \left( z f_{+}\frac{m_{B}}{\sqrt2},0,\textbf{k}_{\rm T}\right),\nonumber\\
p_3&=&\frac{m_{B}}{\sqrt 2}(g_{-},g_{+},\textbf{0}_{\rm T}), ~\quad k_3=\left(0,x_3 g_{+} \frac{m_B}{\sqrt{2}},\textbf{k}_{3{\rm T}}\right),\label{mom-B-k}
\end{eqnarray}
where $m_{B}$ is the $B$ meson mass, and $k_{B}, k$ and $k_3$ are the valence quark momenta in the $B$ meson, the meson pair and the bachelor meson, respectively, with the parton momentum fractions (transverse momenta) $x_B, z$ and $x_3$ (${k}_{B \rm T}, {k}_{\rm T}$ and ${k}_{3{\rm T}}$).
The functions $f_{\pm}$ and $g_{\pm}$ in Eq.~(\ref{mom-B-k}) read
\begin{eqnarray}
f_{\pm}&=&\frac{1}{2}\left(1+\eta-r_3\pm\sqrt{(1-\eta)^2-2r_3(1+\eta)+r_3^2}\right),\nonumber\\
g_{\pm}&=&\frac{1}{2}\left(1-\eta+r_3\pm\sqrt{(1-\eta)^2-2r_3(1+\eta)+r_3^2}\right),\label{fg}
\end{eqnarray}
with the ratios $r_3=m_{P_3}^2/m^2_{B}$ and $\eta=\omega^2/m^2_B$, $m_{P_3}$ being the bachelor meson mass and $\omega^2=p^2$ being the invariant mass squared of the meson pair.
For a $P$-wave meson pair, we introduce the longitudinal polarization vector
\begin{eqnarray}\label{eq:pq1}
\epsilon=\frac{1}{\sqrt{2\eta}}(f_{+},-f_{-},\textbf{0}_{\rm T}).
\end{eqnarray}

The individual meson momenta $p_{1,2}$ in the meson pair are then derived as
\begin{eqnarray}\label{eq:p1p2}
 p_1&=&\left((\zeta+\frac{r_1-r_2}{2\eta})f_{+}\frac{m_B}{\sqrt{2}},
 (1-\zeta+\frac{r_1-r_2}{2\eta})f_{-}\frac{m_B}{\sqrt{2}}, \textbf{p}_{\rm T}\right), \nonumber\\
 p_2&=&\left((1-\zeta-\frac{r_1-r_2}{2\eta})f_{+}\frac{m_B}{\sqrt{2}},
 (\zeta-\frac{r_1-r_2}{2\eta})f_{-}\frac{m_B}{\sqrt{2}}, -\textbf{p}_{\rm T}\right),\nonumber\\
 p_{\rm T}^2&=&\zeta(1-\zeta)\omega^2+\frac{(m_{P_1}^2-m_{P_2}^2)^2}{4\omega^2}-\frac{m^2_{P_1}+m^2_{P_2}}{2},
\end{eqnarray}
from the relation $p=p_1+p_2$ and the on-shell conditions $p_{1,2}^{2}=m_{P_{1,2}}^{2}$, $m_{P_{1,2}}$ being the $P_{1,2}$ meson masses, with the ratios $r_{1,2}=m_{P_1,P_2}^2/m^2_B$. The variable $\zeta+(r_1-r_2)/(2\eta)=p_1^+/p^+$ bears the meaning of the meson momentum fraction up to corrections from the final state masses.
Alternatively, one can define the polar angle $\theta$ of the meson $P_{1}$ in the $P_1P_2$ pair rest frame.
The transformation between the $B$ meson rest frame
and the meson pair rest frame leads to the relation between the meson momentum fraction $\zeta$ and the polar angle $\theta$,
\begin{eqnarray}\label{eq:cos}
2\zeta-1=\sqrt{1-2\frac{r_1+r_2}{\eta}+\frac{(r_1-r_2)^2}{\eta^2}}\cos\theta,
\end{eqnarray}
with the bounds
\begin{eqnarray}
\zeta_{\text{max,min}}=\frac{1}{2}\left[1\pm\sqrt{1-2\frac{r_1+r_2}{\eta}+\frac{(r_1-r_2)^2}{\eta^2}}\right].
\end{eqnarray}

The $B\rightarrow  P_1P_2P_3$ decay amplitude $\mathcal{M}$ is expressed, in the PQCD factorization approach, as
\begin{eqnarray}
\mathcal{M}= \Phi_B \otimes H\otimes \Phi_{P_1P_2} \otimes \Phi_{P_3},
\end{eqnarray}
where $\Phi_B$ ($\Phi_{P_3}$) is the $B$ (bachelor) meson DA, and the two-meson DA $\Phi_{P_1P_2}$ absorbs the nonperturbative dynamics responsible for the production of the meson pair $P_1P_2$.
The hard kernel $H$, containing only one virtual gluon, gathers the perturbative strong and electroweak interactions, as in the formalism for two-body decays.
The symbol $\otimes$ denotes the convolution of the above factors in the parton momenta.
The corresponding branching ratio is given by~\cite{pdg2024}
\begin{eqnarray}\label{eq:br}
\mathcal{B}=\frac{\tau_B m_B}{256\pi^3} \int^1_{(\sqrt{r_1}+\sqrt{r_2})^2} d\eta \sqrt{(1-\eta)^2-2r_3(1+\eta)+r^2_3}\int^{\zeta_{\text{max}}}_{\zeta_{\text{min}}}d\zeta|\mathcal{M}|^2,
\end{eqnarray}
with the $B$ meson lifetime $\tau_B$.
A direct $CP$ asymmetry ${\cal A}_{CP}$ is defined as
\begin{eqnarray}
{\cal A}_{CP}&=&\frac{{\cal B}(\bar B\to \bar f)-{\cal B}( B\to  f)}{{\cal B}(\bar B\to \bar f)+{\cal B}( B\to  f)}.
\end{eqnarray}

For the four-body decay $B \to V_1V_2 \to (P_1P_2)(P_3P_4)$, the kinematics is described in a way similar to that for a three-body decay, which has been specified in our previous work \cite{Rui:2021kbn}, and will not be presented explicitly here.
There are three helicity amplitudes, $A_0$ for the longitudinal polarization, and $A_{\parallel}$ and $A_{\perp}$ for the transverse polarizations with spins being parallel and perpendicular to each other, respectively.
The associated polarization fractions $f_{h}$, $h=0$, $\parallel$ and $\perp$, the  direct $CP$ asymmetry in each polarization component and the overall asymmetry are defined as
\begin{eqnarray}\label{fcpv}
f_{h}=\frac{{\cal B}_h}{{\cal B}_0+{\cal B}_{||}+{\cal B}_\bot},\quad
\mathcal{A}_{CP}^h=\frac{\mathcal{\bar{B}}_h-\mathcal{B}_h}{\mathcal{\bar{B}}_h+\mathcal{B}_h},\quad
\mathcal{A}^{\text{dir}}_{CP}=\frac{\sum_h \mathcal{\bar{B}}_h
-\sum_h \mathcal{B}_h}{\sum_h \mathcal{\bar{B}}_h+\sum_h \mathcal{B}_h},
\end{eqnarray}
respectively, where $\mathcal{\bar{B}}_h$ is the branching ratio of the $CP$-conjugate channel.
It is obvious that the polarization fractions $f_{h}$ obey the normalization $f_0+f_{\parallel}+f_{\perp}=1$.

\subsection{Distribution Amplitudes}\label{sec:22}
We focus on the leading-power contribution from the $B$ meson wave function, which has been widely adopted in the PQCD calculations~\cite{prd63-054008,prd65-014007,epjc28-515,ppnp51-85,prd85-094003,Li:2012md},
\begin{eqnarray}
\Phi_B= \frac{i}{\sqrt{2N_c}} ({ p \hspace{-2.0truemm}/ }_B +m_B) \gamma_5 \phi_B ( x,b), \label{bmeson}
\end{eqnarray}
with the impact parameter $b$ being conjugate to the parton transverse momentum $k_{B \rm T}$.
The $B$ meson DA $\phi_B (x,b)$ is parametrized as
\begin{eqnarray}
\phi_B(x,b)&=& N_B x^2(1-x)^2\mathrm{exp} \left  [ -\frac{m_B^2 x^2}{2 \omega_{B}^2} -\frac{1}{2} (\omega_{B} b)^2\right] ,
 \label{phib}
\end{eqnarray}
where the constant $N_B$ is related to the $B$ meson decay constant $f_B$ through the normalization condition $\int_0^1dx \; \phi_B(x,b=0)=f_B/(2\sqrt{2N_c})$.
The shape parameter takes the values $\omega_B = 0.40$ GeV for $B^+,B^0$ mesons and $\omega_{B_s}=0.48$ GeV~\cite{prd63-054008,plb504-6,prd63-074009,Hua:2020usv} for a $B^0_s$ meson with 10\% variation in the numerical study below.

The light-cone hadronic matrix element for a $B$ meson contains in fact two DAs $\phi_{B}$ and $\bar \phi_{B}$, which are the linear combinations of $\phi_{B}^+$ and $\phi_{B}^-$ defined in the literature~\cite{Grozin:1996pq}, $\phi_{B}=(\phi_{B}^++\phi_{B}^-)/2$ and $\bar \phi_{B}=(\phi_{B}^+-\phi_{B}^-)/2$.
It has been verified that the next-to-power contribution from $\bar \phi_{B}$ is numerically suppressed~\cite{prd65-014007,epjc28-515,prd103-056006} compared with the leading-power one from $\phi_B$.
For instance, the former contribution to the  $B\to\pi$ transition form factor $F_0^{B\to \pi}$ is about 20\% of the latter as shown in~\cite{prd103-056006}.
The higher-twist $B$ meson DAs have been designated systematically in the heavy quark effective theory~\cite{jhep05-022}, which are decomposed according to the twist and conformal spin assignments up to twist 6.
In principle, all the next-to-leading-power sources can be taken into account for a complete analysis. However, the currently available data are not sufficient for their determination.

The light-cone matrix element for a pseudoscalar meson is decomposed, up to twist 3, into~\cite{prd65-014007,epjc28-515}
\begin{eqnarray}
\Phi_{P}\equiv \frac{i}{\sqrt{2N_C}}\gamma_5
                    \left [{ p \hspace{-2.0truemm}/ }_3 \phi_{P}^{A}(x)+m_{03} \phi_{P}^{P}(x)
                    + m_{03} ({ n \hspace{-2.2truemm}/ }
                    { v \hspace{-2.2truemm}/ } - 1)\phi_{P}^{T}(x)\right ],
\end{eqnarray}
with $P=\pi, K$ and the chiral scale $m_{03}$.
The involved DAs $\phi_P^i$ are expanded, in terms of the Gegenbauer polynomials,
\begin{eqnarray}
C_1^{3/2}(t)=3t, ~\quad\quad C_2^{3/2}(t)=\frac{3}{2}(5t^2-1), ~\quad\quad C_4^{3/2}(t)=\frac{15}{8}(1-14t^2+21t^4).
\end{eqnarray}
as
\begin{eqnarray}
 \phi_{\pi}^A(x) &=& \frac{3f_{\pi}}{\sqrt{6}} x(1-x)[ 1 +a_2^{\pi}C_2^{3/2}(2x-1)+a_4^{\pi}C_4^{3/2}(2x-1)], \nonumber\\
 \phi_{\pi}^P(x) &=& \frac{f_{\pi}}{2\sqrt{6}}[1 +a_{2P}^{\pi}C_2^{1/2}(2x-1)], \nonumber\\
 \phi_{\pi}^T(x) &=& \frac{f_{\pi}}{2\sqrt{6}}(1-2x)[1+a_{2T}^{\pi}(10x^2-10x+1)],\nonumber\\
 \phi_{K}^A(x) &=& \frac{3f_{K}}{\sqrt{6}}x(1-x)[1+a_1^{K}C_1^{3/2}(2x-1)+a_2^{K}C_2^{3/2}(2x-1)+a_4^{K}C_4^{3/2}(2x-1)],\nonumber \\
 \phi_{K}^P(x) &=& \frac{f_{K}}{2\sqrt{6}} [1+a_{2P}^{K}C_2^{1/2}(2x-1)],\nonumber \\
 \phi_{K}^T(x) &=& -\frac{f_{K}}{2\sqrt{6}}[C_1^{1/2}(x)+a_{2T}^{K}C_3^{1/2}(x)],
 \label{eq:pkda}
\end{eqnarray}
with the pion (kaon) decay constant $f_\pi$ ($f_K$).
The Gegenbauer coefficients in the above pion and kaon DAs have been derived at the scale 1 GeV in a recent global analysis~\cite{Hua:2020usv} based on the LO PQCD factorization formulas, which are summarized as
\begin{eqnarray}
a_2^{\pi}&=&0.64\pm0.08,\quad a^{\pi}_{4}=-0.41\pm0.10,\quad a^{\pi}_{2P}=1.08\pm0.15,\quad a^{\pi}_{2T}=-0.48\pm0.33,\non
a^{K}_{1}&=&0.33\pm0.08,\quad a^{K}_{2}=0.28\pm0.10, \quad a^{K}_{4}=-0.40\pm0.07,\quad a^{K}_{2P}=0.24, \quad a^{K}_{2T}=0.35.
\label{eq:genpik}
\end{eqnarray}
Note that the twist-3 DAs $\phi_K^P$ and $\phi_K^T$  were not obtained in Ref.~\cite{Hua:2020usv}, but quoted from the sum-rule results~\cite{prd76-074018}.

The $P$-wave two-meson DAs for both longitudinal and transverse polarizations are decomposed, up to twist 3, into~\cite{plb763-29,Rui:2018hls}
\begin{eqnarray}
\Phi_P^{L}(z,\zeta,\omega)&=&\frac{1}{\sqrt{2N_c}} \left [{ \omega \epsilon\hspace{-1.5truemm}/_p  }\phi_P^0(z,\omega^2)+\omega\phi_P^s(z,\omega^2)
+\frac{{p\hspace{-1.5truemm}/}_1{p\hspace{-1.5truemm}/}_2
  -{p\hspace{-1.5truemm}/}_2{p\hspace{-1.5truemm}/}_1}{\omega(2\zeta-1)}\phi_P^t(z,\omega^2) \right ] (2\zeta-1)\;,\label{pwavel}\\
\Phi_P^{T}(z,\zeta,\omega)&=&\frac{1}{\sqrt{2N_c}}
\Big [\gamma_5{\epsilon\hspace{-1.5truemm}/}_{T}{ p \hspace{-1.5truemm}/ } \phi_P^T(z,\omega^2)
+\omega \gamma_5{\epsilon\hspace{-1.5truemm}/}_{T} \phi_P^a(z,\omega^2)+ i\omega\frac{\epsilon^{\mu\nu\rho\sigma}\gamma_{\mu}
\epsilon_{T\nu}p_{\rho}n_{-\sigma}}{p\cdot n_-} \phi_P^v(z,\omega^2) \Big ]\non
&&\cdot \sqrt{\zeta(1-\zeta)+\alpha}\label{pwavet}\;,
\end{eqnarray}
with the kinematic parameter $\alpha=(r_1-r_2)^2/(4\eta^2)-(r_1+r_2)/(2\eta)$.
The DAs $\phi_P^i$  are expanded as
\begin{eqnarray}
\phi_{\pi\pi}^0(z,\omega^2)&=&\frac{3F_{\pi\pi}^{\parallel}(\omega^2)}{\sqrt{2N_c}}z(1-z)\left[1
+a^0_{2\rho}\frac{3}{2}(5t^2-1)\right] \;,\label{eqphi0}\\
\phi_{\pi\pi}^s(z,\omega^2)&=&\frac{3F_{\pi\pi}^{\perp}(\omega^2)}{2\sqrt{2N_c}}t\left[1
+a^s_{2\rho}(10z^2-10z+1)\right]  \;,\label{eqphis} \\
\phi_{\pi\pi}^t(z,\omega^2)&=&\frac{3F_{\pi\pi}^{\perp}(\omega^2)}{2\sqrt{2N_c}}t^2\left[1
+a^t_{2\rho}\frac{3}{2}(5t^2-1)\right]  \;,\label{eqphit}\\
\phi_{\pi\pi}^T(z,\omega^2)&=&\frac{3F_{\pi\pi}^{\perp}(\omega^2)}
{\sqrt{2N_c}}z(1-z)\left[1+a^{T}_{2\rho}\frac{3}{2}(5t^2-1)\right]\;,\label{eqphitt} \\
\phi_{\pi\pi}^a(z,\omega^2)&=&\frac{3F_{\pi\pi}^{\parallel}(\omega^2)}
{4\sqrt{2N_c}}t[1+a_{2\rho}^a(10z^2-10z+1)]\;,\label{eqphia} \\
\phi_{\pi\pi}^v(z,\omega^2)&=&\frac{3F_{\pi\pi}^{\parallel}(\omega^2)}
{8\sqrt{2N_c}}\left[1+t^2+a^v_{2\rho}(3t^2-1)\right]\;,\label{eqphiv}\\
\phi_{K \pi}^0(z,\omega^2)&=&\frac{3F_{K \pi}^{\parallel}(\omega^2)}{\sqrt{2N_c}} z(1-z)\left[1+a_{1K^*}^{||}3t+a_{2K^*}^{||}\frac{3}{2}(5t^2-1)\right]\;,\label{eqphi0kp}\\
\phi_{K \pi}^s(z,\omega^2)&=&\frac{3F_{K \pi}^{\perp}(\omega^2)}{2\sqrt{2N_c}}t\;,\label{eqphiskp}\\
\phi_{K \pi}^t(z,\omega^2)&=&\frac{3F_{K \pi}^{\perp}(\omega^2)}{2\sqrt{2N_c}} t^2\;,\label{eqphitkp}\\
\phi_{K \pi}^T(z,\omega^2)&=&\frac{3F_{K \pi}^{\perp}(\omega^2)}{\sqrt{2N_c}}z(1-z)\left[1+a_{1K^*}^{\perp}3 t+
a_{2K^*}^{\perp}\frac{3}{2}(5 t^2-1)\right]\;,\label{eqphittkp}\\
\phi_{K \pi}^a(z,\omega^2)&=&\frac{3F_{K \pi}^{\parallel}(\omega^2)}{4\sqrt{2N_c}}t\;,\label{eqphiakp}\\
\phi_{K \pi}^v(z,\omega^2)&=&\frac{3F_{K \pi}^{\parallel}(\omega^2)}{8\sqrt{2N_c}}(1+ t^2)\;,\label{eqphivkp}\\
\phi_{KK}^0(z,\omega^2)&=&\frac{3F_{KK}^{\parallel}(\omega^2)}{\sqrt{2N_c}}z(1-z)\left[1
+a^0_{2\phi}\frac{3}{2}(5t^2-1)\right] \;,\label{eqphi0kk}\\
\phi_{KK}^s(z,\omega^2)&=&\frac{3F_{KK}^{\perp}(\omega^2)}{2\sqrt{2N_c}}t  \;,\label{eqphiskk} \\
\phi_{KK}^t(z,\omega^2)&=&\frac{3F_{KK}^{\perp}(\omega^2)}{2\sqrt{2N_c}}t^2  \;,\label{eqphitkk}\\
\phi_{KK}^T(z,\omega^2)&=&\frac{3F_{KK}^{\perp}(\omega^2)}
{\sqrt{2N_c}}z(1-z)\left[1+a^{T}_{2\phi}\frac{3}{2}(5t^2-1)\right]\;,\label{eqphittkk} \\
\phi_{KK}^a(z,\omega^2)&=&\frac{3F_{KK}^{\parallel}(\omega^2)}
{4\sqrt{2N_c}}t\;, \label{eqphiakk}\\
\phi_{KK}^v(z,\omega^2)&=&\frac{3F_{KK}^{\parallel}(\omega^2)}
{8\sqrt{2N_c}}(1+t^2)\;,\label{eqphivkk}
\end{eqnarray}
with the variable $t=1-2z$ and the time-like form factors $F^{\parallel,\perp}(\omega^2)$.
The twist-3 $K\pi$ DAs $\phi_{K\pi}^{a,v}$ and $KK$ DAs $\phi_{KK}^{s,t}$ have been fixed to the asymptotic forms owing to the limited amount of data.
The Gegenbauer coefficients $a^{0(s,t)}_{2\rho}$, $a^{||}_{1(2)K^*}$ and $a^{0(T)}_{2\phi}$ were derived in recent global investigations of three- and four-body $B$ meson decays based on the PQCD formalism~\cite{Li:2021cnd,Yan:2022kck,Yan:2023yvx}.
Note that the moments $a_{2\rho}^{T,a,v}$ in Eqs.~(\ref{eqphitt})-(\ref{eqphiv}) cannot be acquired from present fit, since only the measured transverse polarization fraction $f_\bot(B^0\to \rho^0K^{*0})$ has been reported~\cite{LHCb:2018hsm} so far.
They are thus set to the values $a_{2\rho}^{T}=0.5\pm 0.5, a_{2\rho}^{a}=0.4\pm 0.4$ and $a_{2\rho}^{v}=-0.5\pm 0.5$ deduced in Ref.~\cite{Rui:2018hls}.

According to the Watson theorem~\cite{Watson:1952ji},
elastic rescattering effects in a final-state meson pair can be absorbed into the time-like form factors $F^{\parallel,\perp}(\omega^2)$ as mentioned before.
We employ the RBW line shape for the form factors
\begin{eqnarray}
\label{BRW}
F^{\parallel}_{K\pi(KK)}(\omega^2)&=&\frac{ m_{K^*(\phi)}^2}{m^2_{K^*(\phi)} -\omega^2-im_{K^*(\phi)}\Gamma_{K^*(\phi)}(\omega^2)} \;,
\end{eqnarray}
associated with the narrow intermediate resonances $K^*$ and $\phi$, $m_{K^*(\phi)}$ and $\Gamma_{K^*(\phi)}$ being the pole mass and width , respectively.
The mass-dependent width is written as
\begin{eqnarray}
\label{BRWl}
\Gamma_{K^*(\phi)}(\omega^2)&=&\Gamma_{K^*(\phi)}\left(\frac{m_{K^*(\phi)}}{\omega}\right)\left(\frac{k(\omega)}{k(m_{K^*(\phi)})}\right)^{(2L_R+1)},
\end{eqnarray}
where $k(\omega)$ is the momentum vector of the decay product measured in the resonance rest frame, and $k(m_{K^*(\phi)})$ is the value of $k(\omega)$ at $\omega=m_{K^*(\phi)}$.
Its explicit expression is given by
\begin{eqnarray}
k(\omega)=\frac{\sqrt{\lambda(\omega^2,m_{h_1}^2,m_{h_2}^2)}}{2\omega},
\end{eqnarray}
with the K$\ddot{a}$ll$\acute{e}$n function $\lambda (a,b,c)= a^2+b^2+c^2-2(ab+ac+bc)$.
The orbital angular momenta $L_R=0,1,...$ correspond to the $S,P,...$ partial-wave resonances in the $\pi\pi, K\pi$ and $KK$ systems.

The contribution from a broad $\rho$ resonance is usually parameterized as the GS model~\cite{Gounaris:1968mw} based on the BW function ~\cite{BW-model} in the experimental survey of multi-body hadronic $B$ meson decays.
Taking into account the $\rho$-$\omega$ interference and excited state contributions, we have
the form factor~\cite{prd86-032013,prd95056008,plb763-29}
\begin{eqnarray}
F^\parallel_{P}(\omega^2)= \left [ {\rm GS}_\rho(\omega^2,m_{\rho},\Gamma_{\rho})
\frac{1+c_{\omega} {\rm BW}_{\omega}(\omega^2,m_{\omega},\Gamma_{\omega})}{1+c_{\omega}}
+\Sigma c_j {\rm GS}_j(\omega^2,m_j,\Gamma_j)\right] \left( 1+\Sigma c_j\right)^{-1},
\label{GS}
\end{eqnarray}
where $m_{\rho,\omega,j}$ ($\Gamma_{\rho,\omega,j}$), $j=\rho^{\prime}(1450),\rho^{\prime \prime}(1700)$ and $\rho^{\prime \prime \prime}(2254)$, are the masses (decay widths) of the series of $\rho$ resonances, and $c_{\omega,j}$ are the corresponding weights.
The function ${\rm GS}_\rho(\omega^2,m_{\rho},\Gamma_{\rho})$ reads
\begin{equation}
{\rm GS}_\rho(\omega^2, m_\rho, \Gamma_\rho) =
\frac{m_\rho^2 [ 1 + d(m_\rho) \Gamma_\rho/m_\rho ] }{m_\rho^2 - \omega^2 + f(\omega^2, m_\rho, \Gamma_\rho)
- i m_\rho \Gamma (\omega^2, m_\rho, \Gamma_\rho)},
\end{equation}
with the factors
\begin{eqnarray}
\Gamma (s, m_\rho, \Gamma_\rho) &=& \Gamma_\rho  \frac{s}{m_\rho^2}
\left( \frac{\beta_\pi (s) }{ \beta_\pi (m_\rho^2) } \right) ^3~,\non
d(m) &=& \frac{3}{\pi} \frac{m_\pi^2}{g^2(m^2)} \ln \left( \frac{m+2 g(m^2)}{2 m_\pi} \right)
   + \frac{m}{2\pi  g(m^2)}
   - \frac{m_\pi^2  m}{\pi g^3(m^2)}~,\non
f(s, m, \Gamma) &=& \frac{\Gamma  m^2}{g^3(m^2)} \left[ g^2(s) [ h(s)-h(m^2) ]
+ (m^2-s) g^2(m^2)  h'(m^2)\right]~,\non
g(s) &=& \frac{1}{2} \sqrt{s}  \beta_\pi (s)~,\quad
h(s) = \frac{2}{\pi}  \frac{g(s)}{\sqrt{s}}  \ln \left( \frac{\sqrt{s}+2 g(s)}{2 m_\pi} \right),
\end{eqnarray}
and $\beta_\pi (s) = \sqrt{1 - 4m_\pi^2/s}$.
For the poorly known form factor $F^{\perp}(\omega^2)$, we assume the relation $F^{\perp}(\omega^2)/F^{\parallel}(\omega^2)\approx f_V^T/f_V$, in which $f_V$ and $f_V^{T}$ are the decay constants of the intermediate vector resonance $V$.

\subsection{$P$-wave Parametrization}
The amplitude of the three-body decay $B(p_B)\to P_1(p_1)P_2(p_2)P_3(p_3)$ contains the factorizable contribution
\begin{eqnarray}
\label{amp1}
 \frac{G_F}{\sqrt{2}}\left \langle  P_1(p_1)
 P_2(p_2) \vert J_\mu \vert 0\right \rangle  \left
\langle P_3(p_3)\vert J^\mu \vert B(p_B) \right \rangle,
\end{eqnarray}
with the vector current operators $J_\mu$.
The $B \to P_3$ vector and scalar transition form factors $F^{BP_3}_1(p^2)$ and $F^{BP_3}_0(p^2)$, respectively, are defined via the matrix element
\begin{eqnarray}
\left \langle P_3(p_3)\vert  J^\mu\vert B(p_B)\right \rangle= \left[(p_B+p_3)^\mu-\frac{m^2_B-m^2_{P_3}}{p^2}p^\mu\right]F^{BP_3}_1(p^2)+\frac{m^2_B-m^2_{P_3}}{p^2}p^{\mu}F^{BP_3}_0(p^2).
\end{eqnarray}
Similarly, the matrix element for the transition from vacuum to the $P_1P_2$ meson pair defines the vector (scalar) form factor $F^{P_1P_2}_1(p^2)$ ($F^{P_1P_2}_0(p^2)$),
\begin{eqnarray}
\left \langle  P_1(p_1)P_2(p_2)\vert J_\mu \vert 0\right \rangle &=& \left[(p_1-p_2)_\mu-\frac{m^2_{P_1}-m^2_{P_2}}{p^2}p_\mu\right] F^{P_1P_2}_1(p^2)+\frac{m^2_{P_1}-m^2_{P_2}}{p^2}p_\mu F^{P_1P_2}_0(p^2).\label{45}
\end{eqnarray}
See Eq.~(36) of Ref.~\cite{Ali:1998eb}.

The above subprocess can be described in an alternative way: the quark pair from a hard decay kernel forms the intermediate vector meson $V$, which propagates following the BW factor, and then proceeds with the $V \to P_1P_2$ transition. We write
\begin{eqnarray}
\label{amp2}
\left \langle  P_1(p_1)P_2(p_2)  \vert J_\mu \vert 0\right \rangle &=&\sum_{\lambda=0,\pm1} \left \langle  P_1(p_1)P_2(p_2)  \vert V,\epsilon^\lambda\right \rangle  \frac{1}{m^2_V -s-im_V \Gamma(s)} \left \langle  V, \epsilon^\lambda\vert J_\mu \vert 0\right \rangle \non
&=&\frac{g^{V \to P_1P_2}}{m^2_V -s-im_V\Gamma(s)} \sum_{\lambda=0,\pm1} \epsilon^\lambda\cdot(p_1- p_2) \left \langle  V,\epsilon^\lambda \vert J_\mu \vert 0\right \rangle\non
&=&\frac{g^{V \to P_1P_2}}{m^2_V -s-im_V \Gamma(s)} \sum_{\lambda=0,\pm1} \epsilon^\lambda\cdot(p_1- p_2)\epsilon_{\mu}^\lambda f_Vm_V,
\end{eqnarray}
where $s=p^2=(p_1+p_2)^2$ is the invariant mass squared of the meson pair, $\epsilon^\lambda$ ($m_V$, $f_V$) represents the polarization vector (mass, decay constant) of the resonance $V$, and the coupling strength $g^{V \to P_1P_2}$ can be evaluated from Eq.~(4.17) in Ref.~\cite{Cheng:2022ysn}.

The completeness relation for the summation over the polarization states $\lambda$
\begin{eqnarray}
	\sum_{\lambda=0,\pm1}\epsilon^\lambda_\mu(p)\epsilon^{\lambda*}_\nu(p)=-\left(g_{\mu\nu}-\frac{p_\mu p_\nu}{p^2} \right),
\end{eqnarray}
demands, according to Eqs.~(\ref{45}) and (\ref{amp2}),
\begin{eqnarray}
\label{npr}
\frac{g^{V \to P_1P_2}}{m^2_V -s-im_V\Gamma(s)}\cdot f_V m_V&=& N_P F^{P_1P_2}_1(p^2).
\end{eqnarray}
The coefficient $N_P$ has been introduced to remedy the possible theoretical mismatch between the meson form factors and the properties of the  intermediate $P$-wave resonance.
One can roughly estimate the values of $N_P$ for the $\pi\pi,K\pi$ and $KK$ meson pairs based on Eqs. (\ref{BRW}), (\ref{GS}) and (\ref{npr}),
\begin{eqnarray}
\label{nco}
N_{\pi\pi}\approx 1.00, \quad N_{K\pi}\approx 1.40, \quad N_{KK}\approx 1.20.
\end{eqnarray}
The above three coefficients will be handled as free parameters, determined in our global fit, and then compared with Eq.~(\ref{nco}) in the next section.

\section{Numerical Analysis}

\subsection{Global Fit}
To leviate the precision of the two-meson DAs, we update the fitted Gegenbauer moments by including the additional four-body decays $B \to \rho K^*\to(\pi\pi)(K\pi)$, $B \to K^* K^*\to(K\pi)(K\pi)$  and $B \to \phi K^*\to(KK)(K\pi)$ in the global study.
The moments $a^{\bot}_{1K^*}$ and $a^{\bot}_{2K^*}$ of the transverse $K\pi$ DA in Eq.~(\ref{eqphittkp}) will be derived for the first time.
We specify the inputted masses and widths (in units of GeV)~\cite{pdg2024} in the numerical analysis,
\begin{eqnarray}
\label{para1}
m_{B}&=&5.280, \quad m_{B_s}=5.367, \quad m_b=4.8, \quad ~~~~m_{K^\pm}=0.494,\nonumber\\
 m_{K^0}&=&0.498, \quad m_{\pi^{\pm}}=0.140, \quad m_{\pi^0}=0.135,\quad\Gamma_\rho=0.1496\nonumber\\
\Gamma_{K^*}&=&0.0473, \quad \Gamma_\phi=0.00425.
\end{eqnarray}
The decay constants (in units of GeV) and the $B$ meson lifetimes (in units of ps) take the values~\cite{prd76-074018,prd95056008}
\begin{eqnarray}
\label{para2}
f_B&=&0.21, ~~~\quad f_{B_s}=0.24, ~~~~~~~~~\quad f_{\rho}=0.216 , ~~~~\quad f^T_{\rho}=0.184,\nonumber\\
f_{\phi(1020)}&=&0.215, ~\quad f_{\phi(1020)}^T=0.186, \quad f_{K^*}=0.217, ~\quad f^T_{K^*}=0.185,\nonumber\\
\tau_{B^0}&=&1.519,~\quad \tau_{B^{\pm}}=1.638, ~~~~~~\quad \tau_{B_{s}}=1.512.
\end{eqnarray}
The Wolfenstein parameters in the CKM matrix are set to
$A=0.836\pm0.015, \lambda=0.22453\pm 0.00044$, $\bar{\rho} = 0.122^{+0.018}_{-0.017}$ and $\bar{\eta}= 0.355^{+0.012}_{-0.011}$~\cite{pdg2024} .

The amplitudes $\cal{M}$ for the considered $B$ meson decays can be expanded in terms of the Gegenbauer moments with the two meson DAs in Eqs.~(\ref{eqphi0})-(\ref{eqphivkk}).
Taking the $B\to K^* \phi\to (K\pi)(KK)$, $B\to \rho K^*\to (\pi\pi)(K\pi)$ and $B\to K^* K^*\to (K\pi)(K\pi)$ decays as examples,
we decompose the longitudinal squared amplitudes $|{\cal{M}}_{K^*\phi}^L|^2$, $|{\cal{M}}_{\rho K^*}^L|^2$ and  $|{\cal{M}}_{K^* K^*}^L|^2$
into the linear combinations of the Gegenbauer moments and their products,
\begin{eqnarray}
\label{amfit1}
|{\cal{M}}_{K^*\phi}^L|^2&=&M_{0K^*\phi}^L+a^{0}_{2\phi}M_{1K^*\phi}^L+a^{||}_{1K^*}M_{2K^*\phi}^L+a^{||}_{2K^*}M_{3K^*\phi}^L
+(a^{0}_{2\phi})^2M_{4K^*\phi}^L\non
&+&(a^{||}_{1K^*})^2M_{5K^*\phi}^L+(a^{||}_{2K^*})^2M_{6K^*\phi}^L+(a^{0}_{2\phi}a^{||}_{1K^*})M_{7K^*\phi}^L+(a^{0}_{2\phi}a^{||}_{2K^*})M_{8K^*\phi}^L\non
&+&(a^{||}_{1K^*}a^{||}_{2K^*})M_{9K^*\phi}^L+(a^{0}_{2\phi})^2a^{||}_{1K^*}M_{10K^*\phi}^L+(a^{0}_{2\phi})^2a^{||}_{2K^*}M_{11K^*\phi}^L\non
&+&a^{0}_{2\phi}(a^{||}_{1K^*})^2M_{12K^*\phi}^L+a^{0}_{2\phi}(a^{||}_{2K^*})^2M_{13K^*\phi}^L+(a^{0}_{2\phi}a^{||}_{1K^*}a^{||}_{2K^*})M_{14K^*\phi}^L\non
&+&(a^{0}_{2\phi}a^{||}_{1K^*})^2M_{15K^*\phi}^L+(a^{0}_{2\phi}a^{||}_{2K^*})^2M_{16K^*\phi}^L+(a^{0}_{2\phi})^2a^{||}_{1K^*}a^{||}_{2K^*}M_{17K^*\phi}^L,
\\\non
|{\cal{M}}_{K^* K^*}^L|^2&=&M_{0K^* K^*}^L+ a^{||}_{1K^*} M_{1K^* K^*}^L+ a^{||}_{2K^*} M_{2K^* K^*}^L+
(a^{||}_{1K^*})^2 M_{3K^* K^*}^L+( a^{||}_{2K^*})^2 M_{4K^* K^*}^L\non
&+&(a^{||}_{1K^*}a^{||}_{2K^*} ) M_{5K^* K^*}^L+(a^{||}_{1K^*})^3 M_{6K^* K^*}^L+( a^{||}_{2K^*})^3 M_{7K^* K^*}^L
+(a^{||}_{1K^*})(a^{||}_{2K^*} )^2 M_{8K^* K^*}^L\non
&+&(a^{||}_{1K^*})^2(a^{||}_{2K^*} ) M_{9K^* K^*}^L+(a^{||}_{1K^*})^4 M_{10K^* K^*}^L+(a^{||}_{1K^*})^3(a^{||}_{2K^*} ) M_{11K^* K^*}^L\non
&+&(a^{||}_{1K^*})^2(a^{||}_{2K^*} )^2 M_{12K^* K^*}^L+(a^{||}_{1K^*})(a^{||}_{2K^*} )^3 M_{13K^* K^*}^L+(a^{||}_{2K^*})^4 M_{14K^* K^*}^L,
\\
\non
\label{amfit2}
|{\cal{M}}_{\rho K^*}^L|^2&=&M_{0\rho K^*}^L+a^{0}_{2\rho}M_{1\rho K^*}^L+a^{s}_{2\rho}M_{2\rho K^*}^L+a^{t}_{2\rho}M_{3\rho K^*}^L
+a^{||}_{1K^*}M_{4\rho K^*}^L+a^{||}_{2K^*}M_{5\rho K^*}^L\non
&+&(a^{0}_{2\rho})^2M_{6\rho K^*}^L+(a^{s}_{2\rho})^2M_{7\rho K^*}^L+(a^{t}_{2\rho})^2M_{8\rho K^*}^L
+(a^{||}_{1K^*})^2M_{9\rho K^*}^L+(a^{||}_{2K^*})^2M_{10\rho K^*}^L\non
&+&(a^{0}_{2\rho}a^{s}_{2\rho})M_{11\rho K^*}^L+(a^{0}_{2\rho}a^{t}_{2\rho})M_{12\rho K^*}^L+
(a^{0}_{2\rho}a^{||}_{1K^*})M_{13\rho K^*}^L+(a^{0}_{2\rho}a^{||}_{2K^*})M_{14\rho K^*}^L\non
&+&(a^{s}_{2\rho}a^{t}_{2\rho})M_{15\rho K^*}^L+
(a^{s}_{2\rho}a^{||}_{1K^*})M_{16\rho K^*}^L+(a^{s}_{2\rho}a^{||}_{2K^*})M_{17\rho K^*}^L+
(a^{t}_{2\rho}a^{||}_{1K^*})M_{18\rho K^*}^L\non
&+&(a^{t}_{2\rho}a^{||}_{2K^*})M_{19\rho K^*}^L+(a^{||}_{1K^*}a^{||}_{2K^*})M_{20\rho K^*}^L
+(a^{0}_{2\rho})^2(a^{||}_{1K^*})M_{21\rho K^*}^L+(a^{0}_{2\rho})^2(a^{||}_{2K^*})M_{22\rho K^*}^L\non
&+&(a^{0}_{2\rho}a^{s}_{2\rho}a^{||}_{1K^*})M_{23\rho K^*}^L+(a^{0}_{2\rho}a^{s}_{2\rho}a^{||}_{2K^*})M_{24\rho K^*}^L
+(a^{0}_{2\rho}a^{t}_{2\rho}a^{||}_{1K^*})M_{25\rho K^*}^L+(a^{0}_{2\rho}a^{t}_{2\rho}a^{||}_{2K^*})M_{26\rho K^*}^L\non
&+&(a^{s}_{2\rho})^2(a^{||}_{1K^*})M_{27\rho K^*}^L+(a^{s}_{2\rho})^2(a^{||}_{2K^*})M_{28\rho K^*}^L
+(a^{s}_{2\rho}a^{t}_{2\rho}a^{||}_{1K^*})M_{29\rho K^*}^L+(a^{s}_{2\rho}a^{t}_{2\rho}a^{||}_{2K^*})M_{30\rho K^*}^L\non
&+&(a^{t}_{2\rho})^2(a^{||}_{1K^*})M_{31\rho K^*}^L+(a^{t}_{2\rho})^2(a^{||}_{2K^*})M_{32\rho K^*}^L+(a^{0}_{2\rho})(a^{||}_{1K^*})^2M_{33\rho K^*}^L
+(a^{0}_{2\rho}a^{||}_{1K^*}a^{||}_{2K^*})M_{34\rho K^*}^L\non
&+&(a^{s}_{2\rho})(a^{||}_{1K^*})^2M_{35\rho K^*}^L+(a^{s}_{2\rho}a^{||}_{1K^*}a^{||}_{2K^*})M_{36\rho K^*}^L
+(a^{t}_{2\rho})(a^{||}_{1K^*})^2M_{37\rho K^*}^L+(a^{t}_{2\rho}a^{||}_{1K^*}a^{||}_{2K^*})M_{38\rho K^*}^L\non
&+&(a^{0}_{2\rho})(a^{||}_{2K^*})^2M_{39\rho K^*}^L+(a^{s}_{2\rho})(a^{||}_{2K^*})^2M_{40\rho K^*}^L+(a^{t}_{2\rho})(a^{||}_{2K^*})^2M_{41\rho K^*}^L
+(a^{0}_{2\rho})^2(a^{||}_{1K^*})^2M_{42\rho K^*}^L\non
&+&(a^{0}_{2\rho})^2(a^{||}_{2K^*})^2M_{43\rho K^*}^L+(a^{s}_{2\rho})^2(a^{||}_{1K^*})^2M_{44\rho K^*}^L+(a^{s}_{2\rho})^2(a^{||}_{2K^*})^2M_{45\rho K^*}^L
+(a^{t}_{2\rho})^2(a^{||}_{1K^*})^2M_{46\rho K^*}^L\non
&+&(a^{t}_{2\rho})^2(a^{||}_{2K^*})^2M_{47\rho K^*}^L+(a^{0}_{2\rho})^2(a^{||}_{1K^*}a^{||}_{2K^*})M_{48\rho K^*}^L
+(a^{0}_{2\rho}a^{s}_{2\rho} a^{||}_{1K^*}a^{||}_{2K^*})M_{49\rho K^*}^L\non
&+&(a^{0}_{2\rho}a^{s}_{2\rho})( a^{||}_{1K^*})^2M_{50\rho K^*}^L+(a^{0}_{2\rho}a^{t}_{2\rho})( a^{||}_{1K^*})^2M_{51\rho K^*}^L
+(a^{0}_{2\rho}a^{t}_{2\rho} a^{||}_{1K^*}a^{||}_{2K^*})M_{52\rho K^*}^L\non
&+&(a^{0}_{2\rho}a^{s}_{2\rho})( a^{||}_{2K^*})^2M_{53\rho K^*}^L+(a^{0}_{2\rho}a^{t}_{2\rho})( a^{||}_{2K^*})^2M_{54\rho K^*}^L
+(a^{s}_{2\rho})^2(a^{||}_{1K^*}a^{||}_{2K^*}) M_{55\rho K^*}^L\non
&+&(a^{s}_{2\rho}a^{t}_{2\rho})(a^{||}_{1K^*})^2 M_{56\rho K^*}^L+(a^{s}_{2\rho}a^{t}_{2\rho} a^{||}_{1K^*}a^{||}_{2K^*})  M_{57\rho K^*}^L
+(a^{s}_{2\rho}a^{t}_{2\rho})(a^{||}_{2K^*})^2  M_{58\rho K^*}^L\non
&+&(a^{t}_{2\rho})^2( a^{||}_{1K^*}a^{||}_{2K^*})  M_{59\rho K^*}^L.
\end{eqnarray}
The transverse squared amplitudes $|{\cal{M}}_{K^*\phi}^{N(T)}|^2$ and $|{\cal{M}}_{K^*K^*}^{N(T)}|^2$ are inferred from the above expressions through the  transformations
\begin{eqnarray}
a^{0}_{2\phi}\rightarrow a^{T}_{2\phi},\quad a^{||}_{1K^*}\rightarrow a^{\bot}_{1K^*},\quad a^{||}_{2K^*}\rightarrow a^{\bot}_{2K^*}, \quad
M_{iK^*\phi}^L \rightarrow M_{iK^*\phi}^{N(T)},\quad M_{iK^*K^*}^L \rightarrow M_{iK^*K^*}^{N(T)}.
\end{eqnarray}
We then compute the coefficients $M$, which contain only the Gegenbauer polynomials, to establish the database for the global fit.
The formulas of the squared amplitudes for the
$B\to VP_3\to P_1P_2P_3$ and  $B \to \rho \rho\to (\pi\pi)(\pi\pi)$, $B \to \phi\phi\to (KK)(KK)$ decays can be found in Refs.~\cite{Yan:2022kck,Yan:2023yvx,Li:2021cnd}.

Repeating the procedure in Ref.~\cite{Hua:2020usv},
we determine the Gegenbauer moments in the two-meson DAs by fitting Eqs.~(\ref{amfit1})$-$(\ref{amfit2}) with the Gegenbauer-moment-independent database to the measured branching ratios and polarization fractions of three- and four-body $B$ meson decays.
The standard nonlinear least-$\chi^2$ (lsq) method~\cite{Peter:2020} is employed, in which the $\chi^2$ function is defined for $n$ pieces of experimental data $v_i\pm \delta v_i$ with the errors  $\delta v_i$ and the corresponding fitted values $v^{\rm{th}}_i$ as
\begin{eqnarray} \label{eq:chi}
	\chi^2= \sum_{i=1}^{n}  \Big(\frac {v_i - v^{\rm{th}}_i}{\delta v_i}\Big)^2.
\end{eqnarray}
The inclusion of more data in the fit decreases statistical uncertainties.
The measurements with significance lower than 3$\sigma$, which do not impose stringent constraints, need not be taken into account.
The data of those modes, which are greatly impacted by subleading contributions according to the existent PQCD evaluations \cite{Li:2006jv,Rui:2011dr},
such as $B^0 \to \pi^0\rho^0\to \pi^0(\pi^+\pi^-)$ and $B^0 \to \rho^0\rho^0 \to (\pi^+\pi^-)(\pi^+\pi^-)$, are also excluded, even though they may have higher precision.

\subsection{Gegenbauer Moments of Two-meson DAs}
The Gegenbauer moments of the twist-2 and twist-3 DAs for the $\pi\pi, K\pi$ and $KK$ pairs from a joint fit with \Red{$\chi^2/d.o.f.=1.6$} are presented in Table \ref{tab:gen}, whose errors mainly arise from experimental uncertainties.
The fitted $N_{\pi\pi}=1.05\pm0.04, N_{K\pi}=1.48\pm0.03$ and $N_{KK}=1.22\pm0.03$ are basically consistent with the expected values in Eq.~(\ref{nco}).
The latter two confirm that the discrepancy for the $P$-wave parametrization is remarkable for the $K\pi$ and $KK$ pairs.
The longitudinal $\pi\pi$ DAs are from the fit to the eight pieces of $B\to P(\rho\to)\pi\pi$ data and the six pieces of $B \to \rho \rho, \rho K^*$ data marked by ``$\dagger$" in Tables~\ref{brthree}-\ref{brfour}.
As mentioned before, it is not practical to fit the parameters $a^{T,v,a}_{2\rho}$ in the transverse $\pi\pi$ DAs, because only the measured transverse polarization fraction $f_\bot(B^0\to \rho^0K^{*0})$ is available. These parameters are thus set to the values extracted in Ref.~\cite{Rui:2018hls}.
It is seen that our Gegenbauer moments deviate from the ones of the $\rho(770)$ meson DAs in QCD sum rules~\cite{ball98};
the $\rho$-$\omega$ mixing effect and the contributions from higher $\rho$ resonances with finite widths have been included in the $\pi\pi$ DAs via Eq.~(\ref{GS}), so they need not be the same as the $\rho(770)$ meson DAs.
The added $B^+\to \rho^0 K^{*+}$ and $B^+\to \rho^+ K^{*0}$ data constrain effectively the corresponding Gegenbauer moments, leading to $a^0_{2\rho}=0.39\pm 0.11$, $a^s_{2\rho}=-0.34\pm 0.26$ and $a^t_{2\rho}=-0.13\pm 0.04$, which are distinct from those in Ref.~\cite{Yan:2023yvx}.

The moments $a^{||}_{1K^*}$ and $a^{||}_{2K^*}$ for the twist-2 $K\pi$ DA $\phi^0_{K\pi}$, and $a^{\bot}_{1K^*}$ and $a^{\bot}_{2K^*}$ for the twist-2 DA $\phi^T_{K\pi}$ are governed by the six pieces of $B_{(s)}\to P(K^*\to)K\pi$ data and the twelve pieces of $B_{(s)} \to K^*\rho, K^*K^*, K^*\phi$ data.
We highlight that $a^{\bot}_{1K^*}=0.61\pm 0.21$ and $a^{\bot}_{2K^*}=0.45\pm 0.06$ in the transverse component $\phi_{K\pi}^{T}$ are attained in the global analysis for the first time,
which are controled by the $B_s^0 \to K^{*0}{\bar K}^{*0}$, $B^+ \to K^{*+}\phi$ and $B^0 \to K^{*0}\phi$ data.
The fitted $B^0_s \to K^\pm(K^{*\mp} \to)K\pi$ and $B^0_s \to \KorKbar^0 (\KorKbar\!^{*0} \to)K\pi$ branching ratios deviate from the central values of the data, in contrast with the other channels considered in the fit.
This is not unexpected because of the involved large experimental errors, which do not give tight constraints actually.
As a test, we remove these two modes, and collect the new outcomes in Table~\ref{tab:genn}.
The central values of the Gegenbauer moments change little relative to those in Table~\ref{tab:gen},
indicating that the measured $B^0_s \to K^\pm(K^{*\mp} \to)K\pi$ and $B^0_s \to \KorKbar^0 (\KorKbar\!^{*0} \to)K\pi$ branching ratios indeed have minor impacts on the fit.
Similarly, the obtained Gegenbauer moments for the $K\pi$ DAs vary away from those of the $K^*$ meson DAs in QCD sum rules~\cite{ball98}.
We point out that $a^{||}_{2K^*}=1.13\pm0.32$ in the previous fit~\cite{Li:2021cnd}, being greater than unity, is in fact not favored in view of the convergence of the Gegenbauer expansion.
The refined result $a^{||}_{2K^*}=-0.75\pm 0.08$ in the present work is much smaller for two factors at least: the inclusions of the parameter $N_{K\pi}$
and of the data from the $B_{(s)} \to K^*\rho, K^*K^*, K^*\phi$ decays, among which the latter two dominated by the $B_{(s)} \to K^*$ transition can impose severe constraints on the $K\pi$ DAs.

The results for $a^0_{2\phi}$ and $a^T_{2\phi}$ in the longitudinal and transverse  twist-2 $KK$ DAs $\phi^0_{KK}$ and $\phi^T_{KK}$, respectively, are controled by the two pieces of $B\to K(\phi \to)KK$ data and the nine pieces of $B_{(s)} \to \phi K^*, \phi\phi$ data.
Analogous to the  $\pi\pi$ and $K\pi$ pairs, the above Gegenbauer moments differ from those of the $\phi$ meson DAs in QCD sum rules~\cite{ball98}, and from $a_{2\phi}^{0}=0.40\pm 0.06$ and $a_{2\phi}^{T}=1.48\pm0.07$ in Ref.~\cite{Yan:2022kck}.
It is evident that the convergence of the Gegenbauer expansion for the $K\pi$ and $KK$ DAs has been made more reliable by adding the extra parameters $N_{K\pi}$ and $N_{KK}$.

\begin{figure}[tbp]
\centerline{\epsfxsize=10cm \epsffile{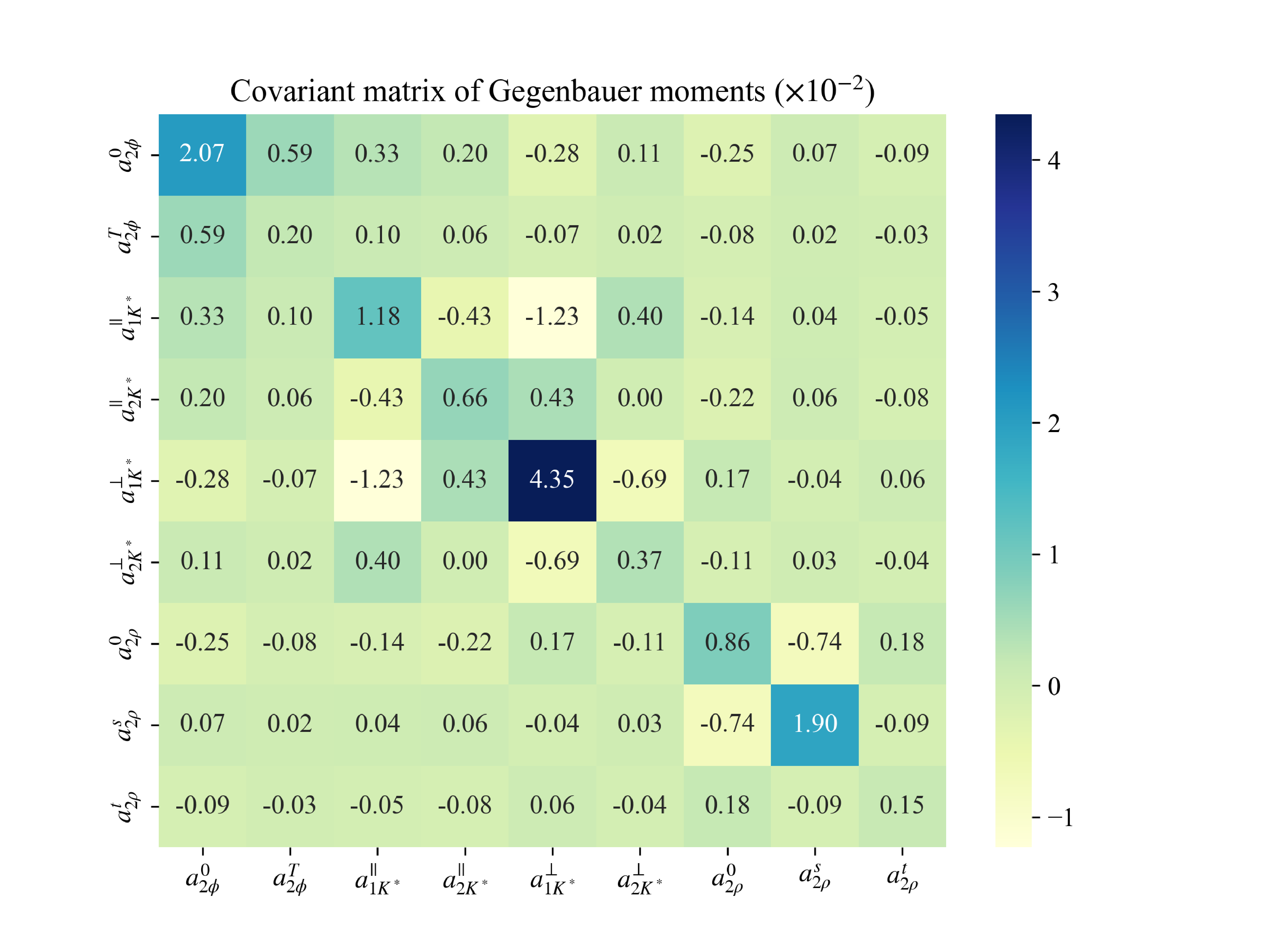}}
\caption{Covariance matrix of the fitted Gegenbauer moments in the twist-2 and twist-3 two-meson DAs. }
\label{cov}
\end{figure}

We remark that the Gegenbauer coefficients fitted from the considered three-body and four-body decays are strongly correlated with each other.
To be more specific,
we present the covariance matrix of the fitted Gegenbauer moments in Fig.~\ref{cov},
which exhibits the correlation between any two parameters via the corresponding matrix element.
For instance, the magnitude of the covariance between $a^{0}_{2\rho}$ and $a^{s}_{2\rho}$ in the twist-2 and twist-3 $\pi\pi$ DAs, respectively, can reach $0.74\times 10^{-2}$,
which is close to the diagonal element $0.86\times 10^{-2}$,
indicating a strong correlation between $a^{0}_{2\rho}$ and $a^{s}_{2\rho}$.
The negative sign means that these two moments tend to vary in an opposite way.

\begin{table}[htbp!]
	\centering
	\caption{Fitted Gegenbauer moments and parameters in the twist-2 and twist-3 two-meson DAs.}
\begin{ruledtabular}
\begin{threeparttable}
    \setlength{\tabcolsep}{1mm}{
	\begin{tabular}{lcccccc}
		
		 &$a^0_{2\rho}$             &$a^s_{2\rho}$                &$a^t_{2\rho}$                         &                &     \\ \hline
\ fit \  &$0.16\pm0.10$                     &$-0.11\pm0.14$            &$-0.21\pm0.04$                   &              &  \\  \hline
&$a_{1K^*}^{||} $            &$a_{2K^*}^{||} $  &$a_{1K^*}^{\bot} $            &$a_{2K^*}^{\bot} $     &   \\ \hline
\ fit \     &$0.45\pm0.11$ & $-0.75\pm0.08$ &$0.61\pm0.21$                                &$0.45\pm0.06$              &  \\  \hline
&$a^0_{2\phi}$     &$a^T_{2\phi}$  &$N_{\pi\pi}$    &$N_{K\pi}$            &$N_{KK}$    &    \\ \hline
\ fit \   &$-0.54\pm0.14$         &$0.77\pm0.04$  &$1.05\pm0.04$ & $1.48\pm0.03$ &$1.22\pm0.03$                           \\

\end{tabular}}
\end{threeparttable}
\end{ruledtabular}
\label{tab:gen}
\end{table}

\begin{table}[htbp!]
	\centering
	\caption{$CP$ averaged branching ratios (in units of $10^{-6}$) of the three-body decays $B\to P_3(V\to P_1P_2)$ in the PQCD approach.
The data for comparison are quoted from~\cite{pdg2024}.
Those data marked by ``$\dagger$" are included in the fit.
The theoretical uncertainties are attributed to the variations of the shape parameter $\omega_{B_{(s)}}$ in the $B_{(s)}$ meson DA, of the Gegenbauer moments in the two-meson DAs and of the hard scale $t$, respectively.}
\begin{ruledtabular}
\begin{threeparttable}
	\begin{tabular}{lcc}
Channels                                     &Results               &Data  \\\hline
$B^+ \to K^+(\rho^0\to)\pi\pi$              &$2.90^{+0.35+0.50+1.38 }_{-0.27-0.31-0.71 }$   &$3.7\pm 0.5$~\tnote{$\dagger$}~~  \\
$B^+ \to K^0(\rho^+\to)\pi\pi$             &$6.77^{+0.89+0.99+3.42 }_{-0.82-0.84-1.94 }$       &$7.3^{+1.0}_{-1.2}$~\tnote{$\dagger$}~~                      \\
$B^0 \to K^+(\rho^-\to)\pi\pi$              &$8.31^{+1.43+1.26+3.89 }_{-1.12-1.59-2.16 }$   &$7.0\pm 0.9$~\tnote{$\dagger$}~~                                  \\
$B^0 \to K^0(\rho^0\to)\pi\pi$             &$3.53^{+0.47+0.48+0.94 }_{-0.37-0.43-0.72 }$   &$3.4\pm 1.1$~\tnote{$\dagger$}~~       \\
$B_s^0 \to K^-(\rho^+\to)\pi\pi$            & $16.9^{+4.8+0.2+0.9 }_{-5.9-0.1-2.4 }$       &$\cdots$                                             \\
$B^0_s \to {\bar K}^0 (\rho^0\to)\pi\pi$       & $0.18^{+0.02+0.02+0.02 }_{-0.02-0.02-0.01 }$             &$\cdots$                                       \\
$B^+ \to \pi^+(\rho^0\to)\pi\pi$                   &$7.05^{+1.72+0.66+0.30 }_{-1.26-0.57-0.35 }$   & $8.3\pm 1.2$~\tnote{$\dagger$}~~                    \\
$B^+ \to \pi^0(\rho^+\to)\pi\pi$                             &$10.07^{+4.20+0.48+0.14 }_{-2.80-0.43-0.10 }$    & $10.6^{+1.2}_{-1.3}$~\tnote{$\dagger$}~~ \\
$B^0 \to \pi^0(\rho^0\to)\pi\pi$              &$0.04^{+0.01+0.01+0.02}_{-0.00-0.01-0.01} $ & $2.0\pm 0.5$          \\
$B^0 \to \pi^{\pm}(\rho^{\mp}\to)\pi\pi$                 &$30.49^{+8.91+2.35+1.26 }_{-6.08-2.08-1.03 }$    & $23.0\pm 2.3$~\tnote{$\dagger$}~~             \\
$B_s^0 \to \pi^+(\rho^-\to)\pi\pi$                            & $0.17^{+0.01+0.04+0.05 }_{-0.01-0.03-0.03 }$     &$\cdots$                                   \\
$B_s^0 \to \pi^-(\rho^+\to)\pi\pi$                              & $0.12^{+0.01+0.02+0.00 }_{-0.01-0.02-0.01 }$    &$\cdots$                                \\
$B_s^0 \to \pi^0(\rho^0\to)\pi\pi$                                    & $0.16^{+0.01+0.02+0.04 }_{-0.01-0.01-0.03 }$      &$\cdots$                         \\
\hline
$B^+ \to K^+({\bar K}^{*0}\to)K\pi$                   &$0.59^{+0.19+0.03+0.12 }_{-0.13-0.02-0.08 }$    & $0.59\pm 0.08$~\tnote{$\dagger$}~~              \\
$B^+ \to {\bar K}^0 (K^{*+}\to)K\pi$                                     & $0.21^{+0.05+0.03+0.07 }_{-0.04-0.03-0.04 }$    &$\cdots$                        \\
$B^0 \to K^\pm(K^{*\mp} \to)K\pi$                                    &$0.51^{+0.01+0.15+0.01}_{-0.02-0.11-0.02} $   &$<0.4$        \\
$B^0 \to \KorKbar^0 (\KorKbar\!^{*0} \to)K\pi$                         &$0.60^{+0.17+0.04+0.12}_{-0.12-0.04-0.08} $ &$<0.96$       \\
$B^0_s \to K^\pm(K^{*\mp} \to)K\pi$                                &$9.35^{+2.11+0.50+1.92 }_{-1.42-0.91-1.24 }$  & $19\pm 5$~\tnote{$\dagger$}~~        \\
$B^0_s \to \KorKbar^0 (\KorKbar\!^{*0} \to)K\pi$                  &$9.31^{+2.28+0.60+2.07 }_{-1.49-0.48-1.34 }$    &$20\pm 6$~\tnote{$\dagger$}~~        \\
$B^+ \to \pi^+(K^{*0}\to)K\pi$                             &$8.89^{+2.60+0.31+2.31 }_{-1.90-0.32-1.60 }$      &$10.1\pm 0.8$~\tnote{$\dagger$}~~         \\
$B^+ \to \pi^0(K^{*+}\to)K\pi$                                       &$5.84^{+1.75+0.23+1.17 }_{-1.27-0.21-0.83 }$  &$6.8\pm 0.9$~\tnote{$\dagger$}~~   \\
$B^0 \to \pi^-(K^{*+}\to)K\pi$                                      &$7.40^{+2.20+0.17+1.60 }_{-1.50-0.12-1.20 }$   &$7.5\pm 0.4$~\tnote{$\dagger$}~~   \\
$B^0 \to \pi^0(K^{*0}\to)K\pi$                                       &$2.80^{+0.76+0.10+0.78 }_{-0.51-0.07-0.53 }$  &$3.3\pm 0.6$~\tnote{$\dagger$}~~   \\
$B_s^0 \to \pi^+(K^{*-}\to)K\pi$                               &$3.63^{+1.50+0.76+0.17}_{-1.01-0.76-0.14} $ &$2.9\pm 1.1$           \\
$B^0_s \to \pi^0 ({\bar K}^{*0}\to)K\pi$                        & $0.11^{+0.02+0.01+0.02 }_{-0.02-0.01-0.01 }$                   &$\cdots$                  \\

\hline
$B^+ \to K^+(\phi\to)KK$                                        &$9.45^{+3.63+0.36+2.91 }_{-2.43-0.22-2.17}$ &$8.8^{+0.7}_{-0.6}$~\tnote{$\dagger$}~~\\
$B^0 \to K^0(\phi\to)KK$                                               &$8.63^{+3.30+0.33+2.66 }_{-2.28-0.24-2.02 }$   &$7.3\pm 0.7$~\tnote{$\dagger$}~~\\
$B^0_s \to {\bar K}^0 (\phi\to)KK$                                    & $0.043^{+0.008+0.013+0.000 }_{-0.006-0.002-0.013 }$      &$\cdots$                        \\
$B^+ \to \pi^+(\phi\to)KK$                                       &$0.011^{+0.004+0.002+0.002}_{-0.003-0.002-0.002} $   &$0.032\pm 0.015$  \\
$B^0 \to \pi^0(\phi\to)KK$                                               &$0.005^{+0.002+0.001+0.001}_{-0.001-0.001-0.001} $  &$<0.015$    \\
$B_s^0 \to \pi^0(\phi\to)KK$                                         & $0.11^{+0.04+0.01+0.01 }_{-0.03-0.01-0.01 }$    &$\cdots$                           \\
\end{tabular}
\end{threeparttable}
\end{ruledtabular}
\label{brthree}
\end{table}

\begin{table}[htbp!]
	\centering
	\caption{$CP$ averaged branching ratios ${\cal B}$ (in units of $10^{-6}$) and polarization fractions (in units of $\%$) of the four-body decays $B\to V_1V_2\to (P_1P_2)(P_3P_4)$ in the PQCD approach, where the final states $(P_1P_2)(P_3P_4)$ are not written for simplicity.
The data for comparison are quoted from~\cite{pdg2024}.
Those data marked by ``$\dagger$" are included in the fit.
The theoretical errors come from the same sources as in Table~\ref{brthree}, but are added in quadrature.}
\begin{ruledtabular}
\begin{threeparttable}
    \setlength{\tabcolsep}{1mm}{
	\begin{tabular}{lcccccccccc}
\multirow{3}{*}{Channels}                &\multicolumn{3}{c}{Results}                       &\multicolumn{3}{c}{Data}\cr\cline{2-7}
	                &${\cal B}(10^{-6}) $      & $f_{0}(\%) $   & $f_{\bot}(\%) $      &${\cal B}(10^{-6}) $       &$f_{0}(\%) $  & $f_{\bot}(\%) $\cr \hline

$B^+ \to \rho^+ \rho^0$         &$12.7^{+4.5}_{-3.3}$  &$97.7^{+0.8}_{-0.9}$  &$1.3^{+0.5}_{-0.4}$          &$24\pm 1.9 $      &$95.0\pm 1.6$     & $\cdots $ \\
$B^0 \to \rho^+\rho^-$          &$27.0^{+10.7}_{-7.5}$  &$92.2^{+3.6}_{-4.3}$  &$4.5^{+2.3}_{-1.9}$           &$27.7\pm 1.9$~\tnote{$\dagger$}~~      &$99.0^{+2.1}_{-1.9}$~\tnote{$\dagger$}~~     & $\cdot\cdot\cdot$\\
$B^0 \to \rho^0 \rho^0$         &$0.35^{+0.12}_{-0.07}$  &$37.9^{+9.4}_{-3.2}$  &$33.9^{+2.6}_{-5.7}$          &$0.96\pm 0.15$  &$71^{+8}_{-9} $  & $\cdots $\\
$B^0_s \to \rho^+ \rho^-$      &$1.35^{+0.84 }_{-0.43 }$ &$99.4^{+0.6 }_{-0.7 }$ & $0.3^{+0.1 }_{-0.2 }$       &$\cdots$&$\cdots$&$\cdots$    \\
$B_s^0 \to \rho^0 \rho^0$          &$0.68^{+0.42}_{-0.22}$  &$99.4^{+0.6}_{-0.7}$  &$0.3^{+0.1}_{-0.2}$        &$<320 $      &$\cdots $     & $\cdots $\\
$B^+ \to \rho^0 K^{*+}$          &$6.92^{+2.06}_{-3.21}$  &$57.8^{+15.1}_{-12.8}$  &$16.9^{+6.2}_{-6.9}$         &$4.6\pm 1.1$~\tnote{$\dagger$}~~      &$78\pm12$~\tnote{$\dagger$}~~     & $\cdot\cdot\cdot$  \\
$B^+ \to \rho^+ K^{*0}$           &$11.1^{+5.81}_{-3.82}$  &$48.7^{+17.5}_{-13.4}$  &$25.7^{+7.6}_{-8.5}$        &$9.2\pm 1.5$~\tnote{$\dagger$}~~      &$48\pm8$~\tnote{$\dagger$}~~     & $\cdot\cdot\cdot$\\
$B^0 \to \rho^- K^{*+}$           &$9.91^{+4.84}_{-3.29}$  &$48.0^{+16.5}_{-12.9}$  &$26.2^{+6.3}_{-8.0}$         &$ 10.3\pm 2.6$      &$38\pm 13 $     & $\cdots $  \\
$B^0 \to \rho^0 K^{*0}$           &$4.35^{+2.40}_{-1.59}$  &$33.2^{+10.9}_{-9.9}$&$41.2^{+4.6}_{-7.8}$           &$3.9\pm 1.3 $      &$ 17.3\pm 2.6$     & $40\pm 4 $\\
$B_s^0 \to \rho^0 {\bar K}^{*0}$  &$0.35^{+0.11}_{-0.06}$  &$59.4^{+9.4}_{-9.6}$  &$21.5^{+4.9}_{-4.8}$          &$ <767$      &$\cdots $     & $\cdots $          \\
$B^0_s \to \rho^+ K^{*-}$         &$12.1^{+ 4.4}_{-3.9 }$ &$89.4^{+1.5 }_{-2.3 }$& $5.3^{+1.2 }_{-0.9 }$       &$\cdots$&$\cdots$&$\cdots$    \\
$B^+  \to \rho^+\phi$            &$0.025^{+0.013}_{-0.009}$  &$87.4^{+4.9}_{-7.2}$  &$5.8^{+2.9}_{-1.9}$           &$<3.0 $      &$\cdots $     & $\cdots $\\
$B^0  \to \rho^0\phi$          &$0.012^{+0.006}_{-0.004}$   &$87.4^{+4.9}_{-7.2}$  &$5.8^{+2.9}_{-1.9}$            &$ <0.33$      &$\cdots $     & $\cdots $\\
$B^0_s \to \rho^0\phi$           &$0.20^{+0.09}_{-0.06}$  &$82.9^{+2.7}_{-1.9}$&$8.9^{+1.1}_{-1.4}$              &$ 0.27\pm 0.08$      &$\cdots $     & $\cdots $ \\
$B^0 \to \phi\phi$           &$0.015^{+0.004}_{-0.003}$  &$98.6^{+0.7}_{-2.0}$  &$0.01^{+0.01}_{-0.00}$                    &$ <0.027$      &$\cdots $     & $\cdots $ \\
$B_s^0 \to \phi\phi$          &$16.6^{+6.6}_{-4.8}$  &$38.7^{+10.6}_{-10.3}$  & $30.9^{+5.1}_{-5.5}$             &$18.5\pm 1.4$~\tnote{$\dagger$}~~      &$37.9\pm 0.8$~\tnote{$\dagger$}~~     & $31.0\pm 0.6$~\tnote{$\dagger$}~~\\
$B^+ \to \phi K^{*+}$         &$11.5^{+4.4}_{-3.9}$  &$54.6^{+4.6}_{-9.1}$  & $23.1^{+4.5}_{-2.2}$             &$10\pm 2$~\tnote{$\dagger$}~~      &$55.0\pm 5.0$~\tnote{$\dagger$}~~     & $20.0 \pm 5.0$~\tnote{$\dagger$}~~\\
$B^0 \to \phi K^{*0}$       &$10.4^{+4.6}_{-3.5}$  &$51.9^{+6.3}_{-8.7}$  & $24.5^{+4.2}_{-3.1}$               &$10.00\pm 0.50$~\tnote{$\dagger$}~~      &$49.7\pm 1.7$~\tnote{$\dagger$}~~     & $22.4\pm 1.5$~\tnote{$\dagger$}~~\\
$B^0_s \to \phi{\bar K}^{*0}$        &$0.29^{+0.17}_{-0.10}$  &$62.3^{+11.9}_{-13.2}$  &$25.2^{+8.8}_{-8.2}$        &$ 1.14\pm 0.30$      &$51\pm 17 $     & $\cdots $  \\
$B^+ \to K^{*+}{\bar K}^{*0}$        &$0.71^{+0.34}_{-0.16}$  &$83.5^{+5.0}_{-3.8}$ &$8.5^{+1.5}_{-3.3}$          &$ 0.91\pm 0.29 $      &$ 82^{+15}_{-21}$     & $\cdots $\\
$B^0 \to K^{*+}{ K}^{*-}$          &$1.24^{+0.38}_{-0.32}$  &$\sim 100$  &$\sim 0$                       &$ <2.0$      &$\cdots $     & $\cdots $ \\
$B^0 \to K^{*0}{\bar K}^{*0}$       &$0.60^{+0.22}_{-0.14}$  &$81.1^{+3.2}_{-5.2}$  &$9.6^{+2.7}_{-1.7}$           &$0.83\pm 0.24$~\tnote{$\dagger$}~~      &$74\pm 5$~\tnote{$\dagger$}~~     & $\cdot\cdot\cdot$   \\
$B^0_s \to K^{*+} K^{*-}$           &$13.7^{+5.4 }_{-3.7 }$ &$32.3^{+10.5 }_{-10.6 }$& $33.9^{+5.3 }_{-5.2 }$     &$\cdots$&$\cdots$&$\cdots$     \\
$B_s^0 \to K^{*0}{\bar K}^{*0}$      &$13.2^{+5.2}_{-3.7}$  &$28.2^{+8.8}_{-9.5}$  & $35.7^{+4.7}_{-4.2}$        &$11.1\pm 2.7$~\tnote{$\dagger$}~~      &$24\pm 4$~\tnote{$\dagger$}~~     & $38\pm 12$~\tnote{$\dagger$}~~  \\
\end{tabular}}
\end{threeparttable}
\end{ruledtabular}
\label{brfour}
\end{table}

\begin{table}[htbp!]
	\centering
	\caption{Fitted Gegenbauer moments and parameters in the twist-2 and twist-3 two-meson DAs without the $B^0_s \to K^\pm(K^{*\mp} \to)K\pi$ and $B^0_s \to \KorKbar^0 (\KorKbar\!^{*0} \to)K\pi$ data.}
\begin{ruledtabular}
\begin{threeparttable}
    \setlength{\tabcolsep}{1mm}{
	\begin{tabular}{lcccccc}
		
		 &$a^0_{2\rho}$             &$a^s_{2\rho}$                &$a^t_{2\rho}$                         &                &     \\ \hline
\ fit \  &$0.18\pm0.10$                     &$-0.11\pm0.14$            &$-0.21\pm0.04$                   &              &  \\  \hline
&$a_{1K^*}^{||} $            &$a_{2K^*}^{||} $  &$a_{1K^*}^{\bot} $            &$a_{2K^*}^{\bot} $     &   \\ \hline
\ fit \     &$0.43\pm0.11$ & $-0.76\pm0.08$ &$0.63\pm0.21$                                &$0.44\pm0.06$              &  \\  \hline
&$a^0_{2\phi}$     &$a^T_{2\phi}$  &$N_{\pi\pi}$    &$N_{K\pi}$            &$N_{KK}$    &    \\ \hline
\ fit \   &$-0.57\pm0.14$         &$0.76\pm0.04$  &$1.05\pm0.04$ & $1.47\pm0.03$ &$1.22\pm0.03$                           \\

\end{tabular}}
\end{threeparttable}
\end{ruledtabular}
\label{tab:genn}
\end{table}

\begin{figure}[tbp]
\centerline{\epsfxsize=9cm \epsffile{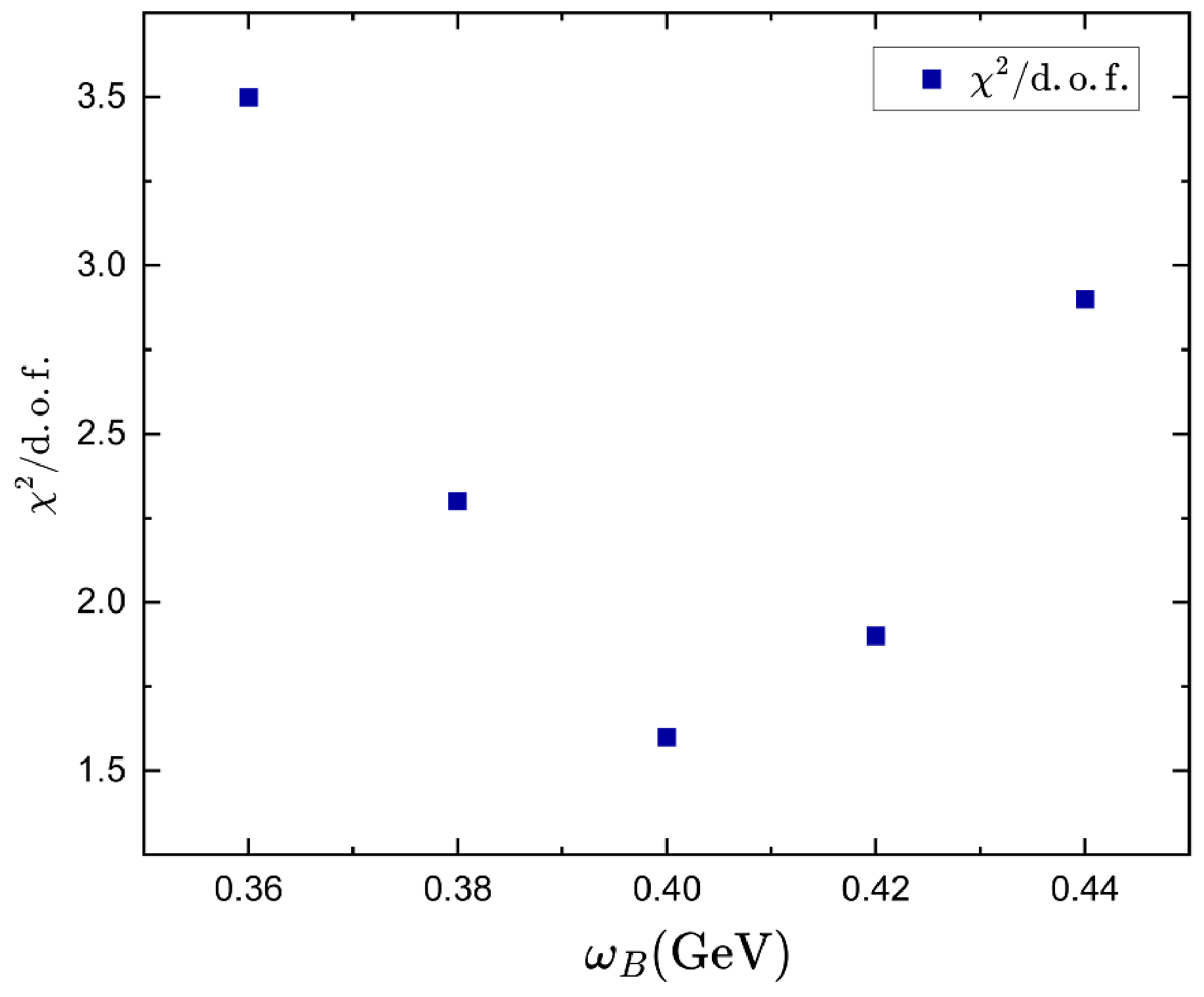}}
\caption{
Dependence of the $\rm \chi^2/d.o.f.$ on the shape parameter $\omega_B$ in the $B$ meson DA. }
\label{chi2}
\end{figure}

\subsection{Branching Ratios of Three-body $B$ Meson Decays}
With the fitted Gegenbauer moments in Table~\ref{tab:gen},
we compute the $CP$ averaged branching ratios of all the considered three-body $B$ meson decays in the LO PQCD formalism, and summarize the predictions in Table~\ref{brthree}.
The listed theoretical uncertainties are assessed from three sources;
the first one is associated with the shape parameter $\omega_B=0.40$~GeV or $\omega_{B_s}=0.48$~GeV with 10\% variation;
the second one originates from the Gegenbauer moments in the two-meson DAs;
the last uncertainty is caused by tuning the hard scale $t$ from $0.75t$ to $1.25t$ to characterize next-to-leading-order (NLO) QCD corrections. The individual uncertainties are then added in quadrature to get the total errors of the LO PQCD predictions in Table~\ref{brthree}.
The errors due to the CKM matrix elements are tiny and can be safely ignored.
Taking three typical decays as examples, we present the  mentioned theoretical uncertainties for the calculated branching ratios ${\cal B}$:
\begin{eqnarray}
{\cal B}(B^+ \to \pi^0(\rho^+\to)\pi\pi)&=&(10.07^{+4.20}_{-2.80}(\omega_B)^{+0.48}_{-0.43}(a_{2\rho})^{+0.14}_{-0.10}(t))\times 10^{-6}
\\
{\cal B}(B^0 \to K^\pm(K^{*\mp} \to)K\pi)&=&(0.51^{+0.01}_{-0.02}(\omega_B)^{+0.15}_{-0.11}(a_{2K^*})^{+0.01}_{-0.02}(t))\times 10^{-6}, \\
{\cal B}(B^+ \to K^+(\rho^0\to)\pi\pi)&=&(2.90^{+0.35}_{-0.27}(\omega_B)^{+0.50}_{-0.31}(a_{2\rho})^{+1.38}_{-0.71}(t))\times 10^{-6},
\end{eqnarray}
where the second errors labeled by $a_{2\rho}$ and $a_{ 2 K^*}$
are derived by adding in quadrature those from the individual Gegenbauer moments in the two-meson DAs.
It is seen that the $B^+ \to \pi^0(\rho^+\to)\pi\pi$ and $B^0 \to K^\pm(K^{*\mp} \to)K\pi$  branching ratios are more sensitive to the variations of the shape parameter $\omega_B$ in the $B$ meson DA and of the Gegenbauer moments in the two-meson DAs, respectively.
For $B^+ \to K^+(\rho^0\to)\pi\pi$ mode, the variation of the hard scale $t$ contributes the main uncertainty.
Generally speaking, the shape parameter $\omega_B$ is the major source of theoretical uncertainties for most exclusive $B$ meson decays~~\cite{Liu:2015sra}.
The combined errors from the three sources could exceed $50\%$, implying that the nonperturbative parameters in the initial and final state DAs need to be known more precisely, and higher-order corrections to multi-body $B$ meson decays are critical.

We have also examined the effects on the global fits
from the variation of the shape parameter $\omega_B$ in the $B$ meson DA around its central value $\omega_B=0.4$ GeV,
while the other parameter $\omega_{B_s}=0.48$ $\rm GeV$ in the $B_s$ meson DA is fixed for simplicity.
The outcomes of $\rm \chi^2/d.o.f.$ for five typical $\omega_B$ values displayed in Fig.~\ref{chi2} shows that $\omega_B=0.40$ GeV indeed leads to the best fit quality with $\chi^2$/d.o.f.=1.6.
This input, together with $\omega_{B_s}=0.48$ GeV, were extracted from the $B_{(s)}$ meson transition form factors in lattice QCD and light-cone sum rules, and have been also verified as a favorable choice by the global analysis of two-body charmless hadronic $B$ meson decays in Ref.~\cite{Hua:2020usv}.
On the other hand, the inverse moment of the $B$ meson DA appears in the QCD factorization approach to exclusive $B$ meson decays.
Its determination has been discussed extensively in the light-cone sum rules for $B$ meson transition form factors,  which are formulated in the heavy quark effective theory~\cite{Wang:2015vgv,Wang:2016qii,Wang:2018wfj}, and has been proposed for lattice QCD studies~\cite{Wang:2019msf,Han:2024min,LatticeParton:2024zko}.

Most of the included $B\to P(\rho\to)\pi\pi$ data in Table~\ref{brthree}, in particular those with higher precision, are well reproduced.
The central value of the fitted $B^0 \to \pi^\pm(\rho^\mp\to)\pi \pi$ branching ratio around $30\times 10^{-6}$ is a bit higher than the data.
As illustrated in Ref.~\cite{Rui:2011dr}, NLO corrections can reduce the LO $B^0 \to \pi^\pm(\rho^\mp\to)\pi \pi$ branching ratio by about $40\%$.
The disparity between the PQCD predictions and the data for three-body $B$ meson decays can be diminished effectively by NLO contributions in principle, whose complete investigation, however, goes beyond the scope of this work.
The observables excluded in the fit are predicted at the LO accuracy, and compared with the data in Table~\ref{brthree}.
The predicted branching ratio of the color-suppressed mode $B^0 \to \pi^0(\rho^0\to)\pi \pi$, which suffers substantial subleading effects, is still below the data, similar to the conclusions in the earlier publication on two-body decays~\cite{Rui:2011dr}.

Overall speaking, the distinction between the updated and previous fits~\cite{Li:2021cnd} is more notable for the $B_s \to P(K^{*}\to)K\pi$ branching ratios than for the $B \to P(K^{*} \to)K\pi$ ones.
Taking the  $B_s^0 \to \pi^+(K^{*-}\to)K\pi$ and $B^0 \to \pi^-(K^{*+}\to)K\pi$ decays as examples, we find that the prediction ${\cal{B}}(B_s^0 \to \pi^+(K^{*-}\to)K\pi)=(3.63^{+1.69}_{-1.27})\times 10^{-6}$ is smaller than
$(12.13^{+4.94}_{-3.85})\times 10^{-6}$ in~\cite{Li:2021cnd} by a factor of $\sim 4$, and becomes closer to the data $(2.9\pm 1.1)\times 10^{-6}$.
For the $B^0 \to \pi^-(K^{*+}\to)K\pi$ decay, our branching ratio is similar to that in \cite{Li:2021cnd}.
The above observations are understandable; the former involves the $B_s \to (K^{*}\to) K\pi$ transition form factor, which is more sensitive to the variation of the Gegenbauer moments in the $K\pi$ DAs.

\begin{figure}[tbp]
\centerline{\epsfxsize=9cm \epsffile{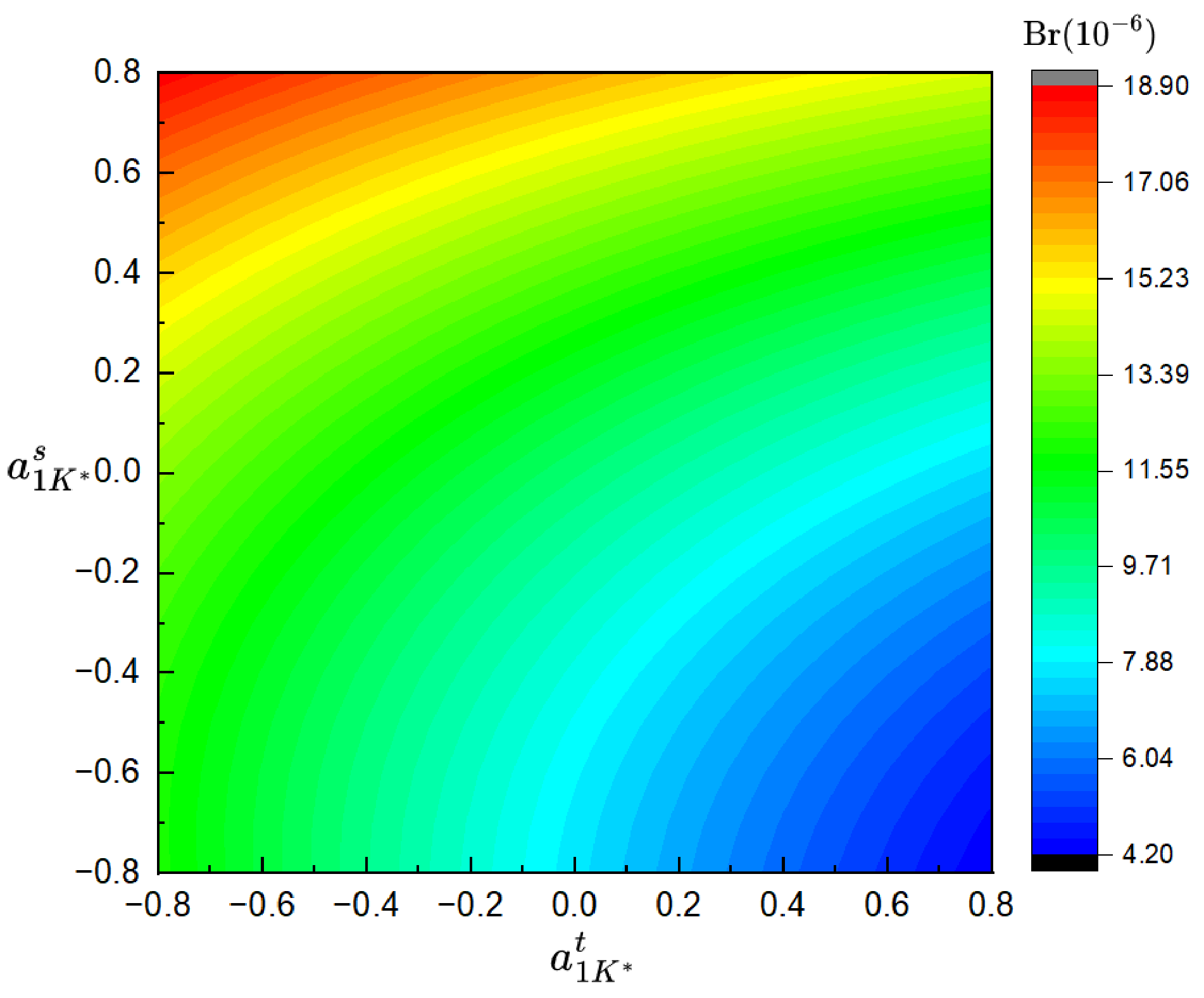}}
\caption{Dependence of ${\cal B}(B^0_s \to K^\pm(K^{*\mp} \to)K\pi)$ on the Gegenbauer moments $a^s_{1K^*}$ and $a^t_{1K^*}$. }
\label{ast}
\end{figure}

As stated before, the predicted branching ratios ${\cal B}(B^0_s \to K^\pm(K^{*\mp} \to)K\pi)$ and ${\cal B}(B^0_s \to \KorKbar^0 (\KorKbar\!^{*0} \to)K\pi)$ are somewhat lower than the data, but consistent with the previous PQCD analysis based on two-body decays~\cite{Yan:2017nlj}.
The authors of Ref.~\cite{Yan:2017nlj} have noticed that the $B^0_s \to K^\pm K^{*\mp}$ and $B^0_s \to \KorKbar^0\KorKbar\!^{*0}$ branching ratios can be enhanced significantly by NLO corrections, which reach $\sim 40\%$ in magnitude of the LO contributions.
We have stuck to the asymptotic forms of the twist-3 $K\pi$ DAs $\phi_{K \pi}^s$ and $\phi_{K \pi}^t$ owing to the finite amount of data in the current study. Here we explore the influence on the above  branching ratios from the subleading Gegenbauer moments in $\phi_{K \pi}^s$ and $\phi_{K \pi}^t$, given by
\begin{eqnarray}
\phi_{K \pi}^s(z,\omega^2)&=&\frac{3F_{K \pi}^{\perp}(\omega^2)}{2\sqrt{2N_c}}[t(1+a_{1K^*}^{s}t)-a_{1K^*}^{s}2z(1-z)]\;,\label{eqphiskpnlo}\\
\phi_{K \pi}^t(z,\omega^2)&=&\frac{3F_{K \pi}^{\perp}(\omega^2)}{2\sqrt{2N_c}} t[t+a_{1K^*}^{t}(3t^2-1)]\;.\label{eqphitkpnlo}
\end{eqnarray}
The dependencies of ${\cal B}(B^0_s \to K^\pm(K^{*\mp} \to)K\pi)$ on the Gegenbauer moments $a^s_{1K^*}$ and $a^t_{1K^*}$ are plotted in Fig.~\ref{ast}, which manifest the sensitivity to $a^s_{1K^*}$ and $a^t_{1K^*}$. It suggests that more precise data of ${\cal B}(B^0_s \to K^\pm(K^{*\mp} \to)K\pi)$ and ${\cal B}(B^0_s \to \KorKbar^0 (\KorKbar\!^{*0} \to)K\pi)$ can impose stronger control on the twist-3 $K\pi$ DAs.

\begin{table}[htbp!]
	\centering
	\caption{Fitted Gegenbauer moments for the twist-2 and twist-3 two-meson DAs with the inclusion of the $a^{s(t)}_{1K^*}$ in the twist-3 $K\pi$ DA $\phi^{s(t)}_{K\pi}$.}
\begin{ruledtabular}
\begin{threeparttable}
    \setlength{\tabcolsep}{1mm}{
	\begin{tabular}{lccccccc}
		
		 &$a^0_{2\rho}$             &$a^s_{2\rho}$                &$a^t_{2\rho}$                         &                &&                    \\ \hline
\ fit \  &$0.09\pm0.12$                     &$-0.08\pm0.14$            &$-0.23\pm0.04$                   &              &&               \\  \hline
&$a_{1K^*}^{||} $            &$a_{2K^*}^{||} $ &$a_{1K^*}^{s} $ &$a_{1K^*}^{t} $ &$a_{1K^*}^{\bot} $            &$a_{2K^*}^{\bot} $     &   \\ \hline
\ fit \     &$0.60\pm0.16$ & $-0.66\pm0.11$& $0.17\pm 0.12$   &$0.02\pm 0.11$ &$0.56\pm0.23$             &$0.46\pm0.07$             \\  \hline
&$a^0_{2\phi}$     &$a^T_{2\phi}$  &$N_{\pi\pi}$    &$N_{K\pi}$            &$N_{KK}$    &  &  \\ \hline
\ fit \   &$-0.57\pm0.15$         &$0.77\pm0.04$  &$1.07\pm0.05$ & $1.45\pm0.04$ &$1.22\pm0.03$   &                        \\

\end{tabular}}
\end{threeparttable}
\end{ruledtabular}
\label{tab:gent3}
\end{table}

\begin{table}[htbp!]
	\centering
	\caption{$CP$ averaged branching ratios (in units of $10^{-6}$) corresponding to the fitted Gegenbauer moments in Table~\ref{tab:gent3} and
compared with the data~\cite{pdg2024}.
 For simplicity, only the theoretical errors
from the Gegenbauer moments are presented.}
\begin{ruledtabular}
\begin{threeparttable}
	\begin{tabular}{lcc}
Channels                                     &Results               &Data  \\\hline
$B^+ \to K^+({\bar K}^{*0}\to)K\pi$                   &$0.58^{+0.04}_{-0.02}$    & $0.59\pm 0.08$              \\
$B^0_s \to K^\pm(K^{*\mp} \to)K\pi$                   &$11.87^{+1.23}_{-0.93}$  & $19\pm 5$         \\
$B^0_s \to \KorKbar^0 (\KorKbar\!^{*0} \to)K\pi$      &$11.79^{+1.31}_{-1.04}$    &$20\pm 6$         \\
$B^+ \to \pi^+(K^{*0}\to)K\pi$                        &$9.07^{+0.32}_{-0.25}$      &$10.1\pm 0.8$         \\
$B^+ \to \pi^0(K^{*+}\to)K\pi$                        &$5.98^{+0.24}_{-0.22}$  &$6.8\pm 0.9$    \\
$B^0 \to \pi^-(K^{*+}\to)K\pi$                        &$7.68^{+0.26}_{-0.20}$   &$7.5\pm 0.4$    \\
$B^0 \to \pi^0(K^{*0}\to)K\pi$                        &$2.86^{+0.11}_{-0.06}$  &$3.3\pm 0.6$    \\
$B_s^0 \to \pi^+(K^{*-}\to)K\pi$                               &$3.85^{+1.17}_{-1.02} $ &$2.9\pm 1.1$           \\
\end{tabular}
\end{threeparttable}
\end{ruledtabular}
\label{kpit3}
\end{table}

We thus perform the global fit for three-body and four-body $B$ meson decays with the inclusion of the subleading Gegenbauer moments $a^s_{1K^*}$ and $a^t_{1K^*}$.
The $B\to PK^*\to PK\pi$ decay amplitude squared $|{\cal A}_{K\pi}|^2$
in Ref.~\cite{Li:2021cnd} is replaced by
\begin{eqnarray}
|{\cal A}_{K\pi}|^2  &=& M_{0K^*}+(a^{0}_{1K^*})M_{1K^*}+(a^{0}_{1K^*})^2M_{2K^*}+a^{0}_{2K^*}M_{3K^*}\\\nonumber
           &+&      (a^{0}_{2K^*})^2M_{4K^*}+      a^{0}_{1K^*}a^{0}_{2K^*}M_{5K^*}+a^{s}_{1K^*} M_{6K^*}+a^{t}_{1K^*}M_{7K^*}\\\nonumber
           &+&(a^{s}_{1K^*})^2M_{8K^*}+(a^{t}_{1K^*})^2M_{9K^*}+(a^{0}_{1K^*}a^{s}_{1K^*}) M_{10K^*}\\\nonumber
           &+&(a^{0}_{2K^*}a^{s}_{1K^*}) M_{11K^*}+
           (a^{0}_{1K^*}a^{t}_{1K^*}) M_{12K^*}\\\nonumber
           &+&(a^{0}_{2K^*}a^{t}_{1K^*}) M_{13K^*}+(a^{s}_{1K^*}a^{t}_{1K^*}) M_{14K^*}.
\end{eqnarray}
The terms associated with $a^{s}_{1K^*}$ and $a^{t}_{1K^*}$ are added into $|{\cal M}^L_{\rho K^*}|^2$, $|{\cal M}^L_{K^* K^*}|^2$ and $|{\cal M}^L_{K^* \phi}|^2$ for the four-body decays in a similar way.
The updated fitting results for all the parameters in the two-meson DAs  with $\chi^2/{\rm d.o.f.}=1.5$,
and for the $B_{(s)}\to P(K^*\to)K \pi$ branching ratios are summarized in Table~\ref{tab:gent3} and Table~\ref{kpit3}, respectively.
As claimed in the previous section, the parameters determined from the multi-body decays reveal strong correlations among each other.
When $a^{s,t}_{1K^*}$  in the twist-3 $K\pi$ DA $\phi^{s,t}_{K\pi}$ are included in the global fit,
the Gegenbauer moments in the $\pi\pi$ and $KK$  DAs change,
but remain consistent with those collected in Table~\ref{tab:gen} within uncertainties.
Relative to $a^t_{1K^*}$, $a^s_{1K^*}$ which gives sizable contributions to the branching ratios, can be constrained more effectively by the data.
We observe that the subleading Gegenbauer moments in twist-3 $K\pi$ DAs enhance the $B^0_s \to K^\pm(K^{*\mp} \to)K\pi$ and $B^0_s \to \KorKbar^0 (\KorKbar\!^{*0} \to)K\pi$ branching ratios by $\sim 30\%$, which reach
${\cal B}(B^0_s \to K^\pm(K^{*\mp} \to)K\pi)=(11.87^{+1.23}_{-0.93})\times 10^{-6}$
and ${\cal B}(B^0_s \to \KorKbar^0 (\KorKbar\!^{*0} \to)K\pi)=(11.79^{+1.31}_{-1.04})\times 10^{-6}$ as shown in Table~\ref{kpit3}.
Though they are still bellow the central values of the data, the consistency between the predictions and measurements has been improved, and is satisfactory in view of the large experimental errors.
Nevertheless, we prefer not to  involve more parameters in the fit due to the limited data, and adhere to the current setup.

With a single Gegenbauer moment $a^0_{2\phi}$ in the twist-2 $KK$ DA $\phi_{KK}^0$, the two pieces of $B\to K(\phi \to)KK$ data included in the fit are well reproduced as indicated in Table~\ref{brthree}.
The predicted $B^+ \to \pi^+(\phi \to)K K$ branching ratio exhibits minor deviation, but still matches the data within uncertainties.

\begin{table}[htbp!]
	\centering
	\caption{PQCD predictions for the $B_{(s)}\to P_1P_2$ transition form factors  $F_\bot(m_R^2, 0)$ with $R=\rho,K^*, \phi$, whose theoretical errors arise from the same sources as in Table~\ref{brthree}, but are added in quadrature.
The results from light-cone sum rules (LCSR)~\cite{Gubernari:2018wyi,Khodjamirian:2006st,Bharucha:2015bzk,Cheng:2017smj,Ball:2004rg,Khodjamirian:2010vf,Descotes-Genon:2019bud} are listed for comparison.
}
\begin{ruledtabular}
\begin{threeparttable}
	\begin{tabular}{lcccccccccccccc}
Decay modes               & This work &LCSR \cite{Gubernari:2018wyi}& LCSR \cite{Khodjamirian:2006st} & LCSR \cite{Bharucha:2015bzk}& LCSR \cite{Cheng:2017smj}& LCSR \cite{Ball:2004rg}& LCSR \cite{Khodjamirian:2010vf}  & LCSR \cite{Descotes-Genon:2019bud}                      \\\hline
$B\to \pi\pi$              &$39^{+9}_{-10}  $  &$29\pm 15$    &$34\pm 11 $   &$35\pm 3$    &$43\pm 12 $    &$34\pm4 $    &$\cdots $  &$\cdots $ \\
$B\to  K\pi$               &$113^{+17}_{-16}  $  &$79\pm 26 $  &$93\pm 26 $     &$81\pm 10 $  &$\cdots $ &$98\pm 11$   &$86\pm 43 $ &$62\pm 36 $     \\
$B_s\to  K\pi$              &$62^{+14}_{-10} $  &$\cdots $  &$\cdots $ &$72\pm 7 $   &$\cdots $ &$75\pm 9 $        &$\cdots $&$\cdots $   \\
$B_s\to  KK$               &$905^{+128}_{-132}$   &$\cdots $  &$\cdots $ &$1080\pm92 $  &$\cdots $ &$ 1121\pm 136$    &$\cdots $ &$\cdots $ \\
\end{tabular}
\end{threeparttable}
\end{ruledtabular}
\label{FFs}
\end{table}

\subsection{$B\to P_1P_2$ Transition Form Factors}

We compute the  $B_{(s)} \to P_1P_2$ transition form factors, which constitute an essential ingredient in the formulation of the semileptonic decays $B_{(s)} \to P_1P_2 l^+l^-$, using the fitted Gegenbauer moments.
The exercise is motivated by the universality of the two-meson DAs in the PQCD factorization.
There are three (seven) form factors involved in the ($S$-wave) $P$-wave $B_{(s)} \to P_1P_2$ transition, whose definitions can be found in Refs.~\cite{Descotes-Genon:2019bud,Cheng:2017smj}.
For simplicity, we focus only on the form factor $F_\bot(\omega^2, Q^2)$ defined via the vector current ${\bar q}\gamma^{\mu} b$, $q=u,d,s$, with the invariant mass $\omega$ of an intermediate resonance $R$ and the momentum transfer $Q$,
\begin{eqnarray}
i\langle P_1(p_1)P_2(p_2)|{\bar q}\gamma^{\mu} b|B(p_B)\rangle&=&F_\bot(\omega^2, Q^2)k^\mu_\bot,
\end{eqnarray}
where the vector $k^\mu_\bot$ has been specified in Ref.~\cite{Descotes-Genon:2019bud}.
The PQCD factorization formula for $F_\bot(\omega^2, Q^2)$ is referred to Eq.~(A.59) in Ref.~\cite{Rui:2021kbn}.
The predictions for $F_\bot(m^2_R, 0)$ with $R=\rho, K^*, \phi$ at maximal recoil are presented in Table.~\ref{FFs}, and compared with those from light-cone sum rules in the narrow-width limit.
The consistency between the two approaches is satisfactory within theoretical uncertainties, hinting that the PQCD formalism and the predictions from it are reasonable.
A complete and systematic PQCD study of all the $B_{(s)}\to P_1P_2$ form factors, including their $\omega^2$ and $Q^2$ dependencies, will be postponed to a future work.

The $B,B_{s} \to K\pi$ transition form factors $F_\bot(m^2_{K^*}, 0)$ have been calculated in the PQCD approach~\cite{Rui:2021kbn}.
We find that the updated $F_\bot(m^2_{K^*}, 0)=113$ for $B^0 \to K\pi$ is larger than the previous one 88, while
the updated $F_\bot(m^2_{K^*},  0)=62$ is very close to the previous one 64 for $B^0_s \to K\pi$.
It is easy to see that the destruction between the contributions from the Gegenbauer polynomials  $C_1^{3/2}(t)$ and $C_2^{3/2}(t)$ in the transverse twist-2  $K\pi$ DA $\Phi^T_{K\pi}$ weakens the dependence of $F_\bot(m^2_{K^*}, 0)$ on the moments $a^\bot_{1K*}$ and $a^\bot_{2K*}$ in the latter case.
For the $B^0 \to K\pi$ transition, the value of $a^\bot_{1{ K}^*}$, related to $SU(3)$ symmetry breaking effects, flips sign,
since the momentum fraction is associated with the ${\bar s}$ quark in the ${\bar K}^{*0}$ meson.
The contributions from the aforementioned $C_1^{3/2}(t)$ and $C_2^{3/2}(t)$ then become constructive, and the associated $F_\bot(m^2_{K^*}, 0)$ is more sensitive to the change of the Gegenbauer moments.

\subsection{Branching Ratios and Polarization Fractions of $B\to V_1V_2$ Decays}
Most of the predicted branching ratios of the four-body decays $B \rightarrow  V_1V_2\rightarrow (P_1P_2) (P_3P_4)$
agree with the data within sizable theoretical errors as evinced in Table~\ref{brfour}.
Our predictions ${\cal B}(B^+\to \rho^0\rho^+)=(12.7^{+4.5}_{-3.3})\times 10^{-6}$ and ${\cal B}(B^0\to \rho^0\rho^0)=(0.35^{+0.12}_{-0.07})\times 10^{-6}$ are slightly below those from other theoretical approaches~\cite{Cheng:2009mu,Wang:2017rmh,Wang:2017hxe} and the data~\cite{pdg2024}, but close to the earlier PQCD results from the two-body-decay framework~\cite{prd91-054033}.
It has been observed that strong cancellations occurs to the nonfactorizable contributions in the color-suppressed mode $B^0\to \rho^0\rho^0$ at LO, and its branching ratio turns out to be as tiny as ${\cal B}(B^0\to \rho^0\rho^0)=(0.35^{+0.10}_{-0.06})\times 10^{-6}$.
Then the relation of ${\cal B}(B^0\to \rho^+\rho^-)\approx2{\cal B}(B^+\to \rho^+\rho^0)$ always holds in the PQCD analysis, because of the small ${\cal B}(B^0\to \rho^0\rho^0)\sim 10^{-7}$ in the isospin triangle.
On the experimental side, nevertheless, the two rates are roughly equal to each other within errors, leading to a puzzle that has persisted for a long time.

It is naively expected that the longitudinal components  dominate the charmless $B\to V_1V_2$ decays based on the counting rules for the polarization fractions~\cite{plb622-63},
\begin{eqnarray}\label{eq:cr}
f_0\sim 1-\mathcal{O}(m_V^2/m_B^2),\quad f_{\parallel}\sim f_{\perp}\sim \mathcal{O}(m_V^2/m_B^2),
\end{eqnarray}
with a vector meson mass $m_V$.
The penguin-dominated modes, such as $B\to K^*\rho$, $B\to K^*\phi$ and $B^0_s\to \phi\phi$ with the measured longitudinal polarization fractions being around $50\%$~\cite{prd85-072005,LHCb:2018hsm,prl91-201801,prd78-092008,prl107-261802,plb713-369}, do not respect the above counting rules.
To interpret the large transverse polarizations, a number of resolutions within or beyond the Standard Model have been proposed~\cite{npb774-64,prd71-054025,prd70-054015,Cheng:2008gxa,Grossman:2003qi,Das:2004hq,Chen:2005mka,Yang:2004pm,Kagan:2004uw,Beneke:2005we,Datta:2007qb,Colangelo:2004rd,Ladisa:2004bp,Cheng:2004ru,
Bobeth:2014rra,Cheng:2010yd,Chen:2007qj,Chen:2005cx,Faessler:2007br,Chen:2006vs,Huang:2005qb,Baek:2005jk,Yang:2005tv,Alvarez:2004ci}.
In the PQCD approach, the substantial transverse contributions are accommodated by means of the hard emission diagrams and the chirally enhanced annihilation diagrams, especially the $(S-P)(S+P)$ penguin annihilation from the QCD penguin operator $O_6$~\cite{prd71-054025,Kagan:2004uw}.

\begin{figure}[tbp]
\centerline{\epsfxsize=9cm \epsffile{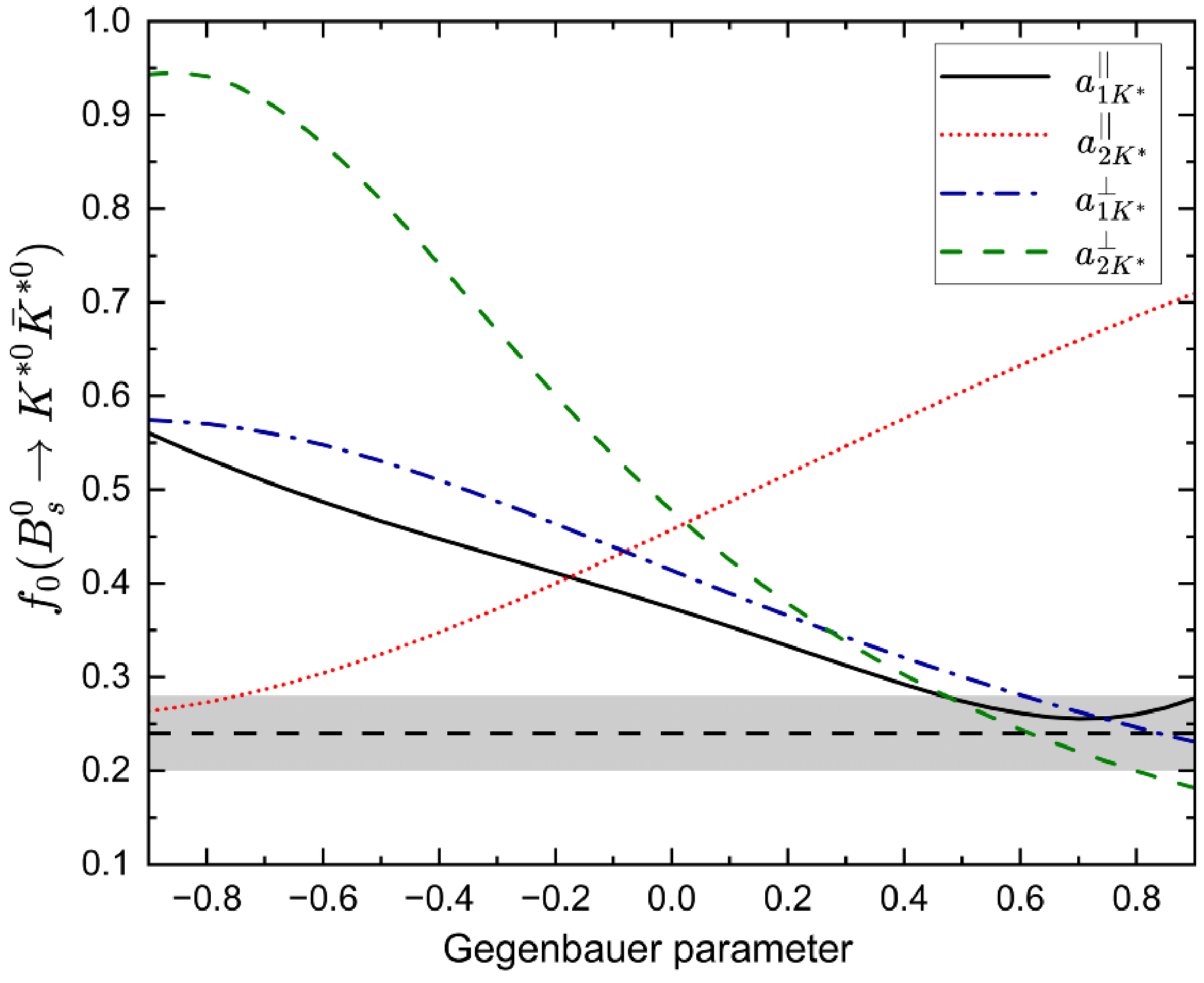}}
\caption{Longitudinal polarization fraction $f_0$ in the $B_s^0\to K^{*0} {\bar K}^{*0}$ decay as a function of the Gegenbauer moments $a^{||(\bot)}_{1K^*}$ and  $a^{||(\bot)}_{2K^*}$.
The dashed horizontal line corresponds to the central value of the data~\cite{pdg2024} with the shaded band representing the experimental errors.}
\label{figlkk}
\end{figure}

The longitudinal polarization fraction for the color-suppressed $B^0\to \rho^0\rho^0$ decay is predicted to be $f_0=(37.9^{+9.4}_{-3.2})\%$, which is much lower than the data $(71^{+8}_{-9})\%$~\cite{pdg2024}.
It has been found that the longitudinal polarization contributions from the hard emission diagrams largely cancel each other in this type of decays.
The minor penguin annihilation contributions then play a role in enhancing the transverse components, such that the transverse polarization fraction $f_\bot$ becomes comparable with the longitudinal one.
We point out that NLO corrections to $f_0(B^0\to \rho^0\rho^0)$ can shrink the gap between the PQCD prediction and the data~\cite{Li:2006cva}, which will be addressed in a future work.
It is worth mentioning that the predicted longitudinal polarization fraction of the pure-penguin decay $f_0(B_s^0\to K^{*0} {\bar K}^{*0})=(28.2^{+8.8}_{-9.5})\%$, far below our previous result~\cite{Rui:2021kbn}, is now in good agreement with the data $f_0(B_s^0\to K^{*0} {\bar K}^{*0})_{\rm exp}=(24\pm 4)\%$~\cite{pdg2024}.
The improved consistency is mainly attributed to the updated Gegenbauer moments $a^{||(\bot)}_{1K^*}$ and $a^{||(\bot)}_{2K^*}$.
To be specific, we display $f_0(B_s^0\to K^{*0} {\bar K}^{*0})$ as a function of the above moments in Fig.~\ref{figlkk}, which decreases quickly as  $a^{||}_{2K^*}$ turns negative and as the other three increase. It is then not hard to realize the drop of $f_0(B_s^0\to K^{*0} {\bar K}^{*0})$ by comparing $a_{1K^*}^{\parallel}=a_{1K^*}^{\perp}=0.31\pm 0.16$ and $a_{2K^*}^{\parallel}=a_{2K^*}^{\perp}=1.188\pm0.099$ employed in~\cite{Rui:2021kbn} with the corresponding ones in Table~\ref{tab:gen}.

It is well known that flavor-changing neutral-current  transitions are very sensitive to new physics (NP) contributions.
Though the polarization anomaly associated with the $B_s^0\to K^{*0} {\bar K}^{*0}$ decay is resolved here, we cannot rule out the possibility of NP effects.
A new observable $L_{K^{*0}{\bar K}^{*0}}$ in terms of two $U$-spin related decays $B_s^0\to K^{*0} {\bar K}^{*0}$ and $B^0\to K^{*0} {\bar K}^{*0}$ has been constructed to maximize the sensitivity to NP by reducing hadronic uncertainties~\cite{Alguero:2020xca},
\begin{eqnarray*}
L_{K^{*0}{\bar K}^{*0}}=\frac{{\cal B}(B_s^0\to K^{*0} {\bar K}^{*0})g(B^0\to K^{*0} {\bar K}^{*0})f_L(B_s^0\to K^{*0} {\bar K}^{*0})}
   {{\cal B}(B^0\to K^{*0} {\bar K}^{*0})g(B_s^0\to K^{*0} {\bar K}^{*0})f_L(B^0\to K^{*0} {\bar K}^{*0})},
\end{eqnarray*}
with the phase space factors
\begin{eqnarray}
g(B_{(s)}\to K^{*0} {\bar K}^{*0} )=\frac{\tau_{B_{(s)}}}{16\pi m^2_{B_{(s)}}} \sqrt{m^2_{B_{(s)}}-4m^2_{K^{*0}}}.
\end{eqnarray}
Adopting the latest inputs from the PDG~\cite{pdg2024}, one  derives the experimental result~\cite{Alguero:2020xca}
\begin{eqnarray}
L^{\rm exp}_{K^{*0}{\bar K}^{*0}}=4.43 \pm 0.92,\label{63}
\end{eqnarray}
smaller than the QCDF~\cite{Alguero:2020xca} and PQCD~\cite{Li:2022mtc} predictions
\begin{eqnarray}
L_{K^{*0}{\bar K}^{*0}}&=&\left\{\begin{array}{ll}
19.5^{+9.3}_{-6.8} &(\text{QCDF}), \\
12.7^{+5.6}_{-3.2} &(\text{PQCD}).\\
\end{array} \right.\label{64}
\end{eqnarray}
The difference between Eqs.~(\ref{63}) and (\ref{64}), being regarded as a NP signal~\cite{Alguero:2020xca}, motivated the authors of Ref.~\cite{Li:2022mtc} to consider a family non-universal $Z^{\prime}$ boson in the $b\to sq {\bar q}$ transition.
We update the ratio $L_{K^{*0}{\bar K}^{*0}}$ based on Table~\ref{brfour},
\begin{eqnarray}
L^{\rm PQCD}_{K^{*0}{\bar K}^{*0}}=7.7^{+4.9}_{-3.8},
\end{eqnarray}
which turns closer to Eq.~(\ref{63}).
Namely, the deviation of our prediction from the data can be alleviated by improving the precision of nonperturbative parameters.
We encourage experimentalists to conduct direct measurements on $L_{K^{*0}{\bar K}^{*0}}$ to uncover potential NP signals.

\subsection{Direct $CP$ Asymmetries of Multi-body $B$ Decays}
\begin{table}[t]
\caption{Direct $CP$ asymmetries (in units of $\%$) of the three-body decays $B\to VP_3\to P_1P_2P_3$.
The data for comparison are quoted from~\cite{pdg2024}.
The theoretical errors come from the same sources as in Table~\ref{brthree}, but are added in quadrature.}
\label{cpthree}
\begin{ruledtabular}
\setlength{\tabcolsep}{1mm}{
\begin{tabular}[t]{lcc }
Modes  & {\rm Results} & Data \\
\hline
$B^+ \to K^+(\rho^0\to)\pi \pi$      &$62.4^{+11.6}_{-13.5}$  &$16\pm 2$ \\
$B^+ \to K^0(\rho^+\to)\pi \pi$     &$9.3^{+1.2}_{-2.2}$  &$-3\pm 15$ \\
$B^0 \to K^+(\rho^-\to)\pi \pi$          &$44.1^{+7.2}_{-7.3}$  &$20\pm 11$ \\
$B^0 \to K^0(\rho^0\to)\pi \pi$     &$1.5^{+2.4}_{-2.2}$  &$-4\pm 20$ \\
$B_s^0 \to K^-(\rho^+\to)\pi \pi$     &$19.6^{+4.4}_{-5.2}$  &$\cdot\cdot\cdot$ \\
$B_s^0 \to \bar K^0(\rho^0\to)\pi \pi$     &$-36.4^{+26.8}_{-17.9}$  &$\cdot\cdot\cdot$ \\
$B^+ \to \pi^+(\rho^0\to)\pi \pi$     &$-34.1^{+7.3}_{-9.8}$  &$0.9\pm 1.9$ \\
$B^+ \to \pi^0(\rho^+\to)\pi \pi$     &$23.1^{+7.0}_{-6.1}$  &$2\pm 11$ \\
$B^0 \to \pi^+(\rho^-\to)\pi \pi$     &$-26.5^{+5.4}_{-5.7}$  &$-8\pm 8$ \\
$B^0 \to \pi^-(\rho^+\to)\pi \pi$     &$9.5^{+2.3}_{-2.6}$  &$13\pm 6$ \\
$B^0 \to \pi^0(\rho^0\to)\pi \pi$     &$16.7^{+23.5}_{-18.9}$  &$27\pm 24$ \\
$B_s^0 \to \pi^+(\rho^-\to)\pi \pi$     &$-37.9^{+19.1}_{-13.1}$  &$\cdot\cdot\cdot$ \\
$B_s^0 \to \pi^-(\rho^+ \to)\pi \pi$     &$-66.2^{+7.0}_{-4.5}$  &$\cdot\cdot\cdot$ \\
$B_s^0 \to \pi^0(\rho^0 \to)\pi \pi$     &$-40.0^{+8.8}_{-4.4}$  &$\cdot\cdot\cdot$ \\

$B^+ \to K^+(\bar{K}^{*0} \to)K\pi$     &$-4.0^{+12.2}_{-10.3}$  &$4\pm 5$ \\
$B^+ \to \bar{K}^0(K^{*+}\to)K \pi$     &$-61.9^{+4.2}_{-19.2}$  &$\cdot\cdot\cdot$ \\
$B^0 \to K^+(K^{*-} \to)K\pi$     &$18.4^{+8.9}_{-6.3}$  &$\cdot\cdot\cdot$ \\
$B^0 \to K^-(K^{*+} \to)K\pi$     &$14.3^{+4.6}_{-9.9}$  &$\cdot\cdot\cdot$ \\
$B_s^0 \to K^+(K^{*-}\to)K\pi$     &$11.5^{+7.2}_{-8.8}$  &$\cdot\cdot\cdot$ \\
$B_s^0 \to K^-(K^{*+}\to)K \pi$     &$-7.3^{+8.0}_{-7.9}$  &$\cdot\cdot\cdot$ \\
$B^+ \to \pi^+(K^{*0}\to)K \pi$     &$-2.3^{+1.2}_{-0.5}$  &$-4\pm 9$ \\
$B^+ \to \pi^0(K^{*+}\to)K \pi$     &$-3.5^{+4.0}_{-3.9}$  &$-39\pm 21$ \\
$B^0 \to \pi^-(K^{*+}\to)K \pi$     &$-15.6^{+5.9}_{-6.4}$  &$-27\pm 4$ \\
$B^0 \to \pi^0(K^{*0}\to)K \pi$     &$-13.5^{+1.6}_{-1.8}$  &$-15\pm 13$ \\
$B_s^0 \to \pi^+(K^{*-}\to)K\pi$     &$-18.1^{+5.7}_{-5.9}$  &$\cdot\cdot\cdot$ \\
$B_s^0 \to \pi^0(\bar{K}^{*0}\to)K \pi$     &$-22.3^{+15.7}_{-17.8}$  &$\cdot\cdot\cdot$ \\

$B^+ \to K^+(\phi \to)KK$     &$0.7^{+1.2}_{-1.8}$  &$2.4\pm 2.8$ \\
$B_s^0 \to \pi^0(\phi \to)K K$     &$31.2^{+3.7}_{-3.7}$  &$\cdot\cdot\cdot$ \\
\end{tabular}}
\end{ruledtabular}
\end{table}

\begin{table}[t]
\caption{Direct $CP$ asymmetries (in units of $\%$) of the four-body decays $B\to V_1V_2\to (P_1P_2)(P_3P_4)$ in the PQCD approach.
The data for comparison are quoted from~\cite{pdg2024}.
The theoretical errors arise from the same sources as in Table~\ref{brthree}, but are added in quadrature.
The percentage in the parentheses specifies the
proportion of CPV of each helicity state to the direct CPV.}
\label{cpfour}
\begin{ruledtabular}
\setlength{\tabcolsep}{1mm}{
\begin{tabular}[t]{lcccc}
Modes  & $\mathcal{A}^0_{\text{CP}}$ & $\mathcal{A}^{\parallel}_{\text{CP}}$& $\mathcal{A}^{\perp}_{\text{CP}}$ & $\mathcal{A}^{\text{dir}}_{\text{CP}}$\\
\hline
$B^+ \to \rho^+ \rho^0\to(\pi^+\pi^0)(\pi^+\pi^-)$  &$ 0.4^{+0.2}_{-0.1}(97.7\%)$&$-0.1 ^{+0.4}_{-0.6}(1.0\%)$ &$0.5 ^{+0.2}_{-0.6}(1.3\%)$&$ 0.4^{+0.2}_{-0.1}$\\
Data   & $\cdots$  &$\cdots$  &$\cdots$ &$-5\pm 5$\\
$B^0 \to \rho^+\rho^-\to(\pi^+\pi^0)(\pi^-\pi^0)$ &$ -3.7^{+0.9}_{-2.0}(92.2\%)$&$ 43.4^{+11.7}_{-17.7}(3.3\%)$ &$38.4 ^{+12.0}_{-16.9}(4.5\%)$&$ -0.3^{+2.9}_{-2.4}$\\
Data   & $\cdots$  &$\cdots$  &$\cdots$ &$0\pm 9$\\
$B^0 \to \rho^0 \rho^0\to(\pi^+\pi^-)(\pi^+\pi^-)$ &$ 61.7^{+12.3}_{-14.8}(37.9\%)$&$64.9 ^{+6.1}_{-8.4}$(28.2\%) &$ 76.9^{+8.8}_{-9.0}(33.9\%)$&$ 67.8^{+7.1}_{-9.9}$\\
Data   & $\cdots$  &$\cdots$  &$\cdots$ &$20\pm 90$\\
$B_s^0 \to \rho^+\rho^-\to(\pi^+\pi^0)(\pi^-\pi^0)$ &$ 5.1^{+4.3}_{-2.5}(99.4\%)$&$ 4.3^{+4.6}_{-3.9}(0.3\%)$ &$ 5.9^{+8.6}_{-5.3}(0.3\%)$&$5.1 ^{+4.2}_{-2.6}$\\
$B_s^0 \to \rho^0 \rho^0\to(\pi^+\pi^-)(\pi^+\pi^-)$ &$ 5.1^{+4.3}_{-2.5}(99.4\%)$&$ 4.3^{+4.6}_{-3.9}(0.3\%)$ &$ 5.9^{+8.6}_{-5.3}(0.3\%)$&$5.1 ^{+4.2}_{-2.6}$\\
$B^+\to\rho^+K^{*0}\to(\pi^+\pi^0)(K^+\pi^-)$ &$ -0.2^{+1.9}_{-3.4}(48.7\%)$&$1.6 ^{+0.9}_{-1.3}(25.6\%)$ &$1.7 ^{+0.2}_{-0.6}(25.7\%)$&$ 0.8^{+0.8}_{-1.8}$\\
Data   & $\cdots$  &$\cdots$  &$\cdots$ &$-1\pm 16$\\
$B^+ \to\rho^0K^{*+}\to(\pi^+\pi^-)(K^0\pi^+)$ &$ 32.8^{+2.7}_{-8.5}(57.8\%)$&$0.8 ^{+4.4}_{-4.6}(25.3\%)$ &$-56.2 ^{+9.7}_{-9.4}(16.9\%)$&$9.6 ^{+7.1}_{-7.7}$\\
Data  & $\cdots$  &$\cdots$  &$\cdots$ &$31\pm 13$\\
$B^0 \to \rho^0K^{*0}\to(\pi^+\pi^-)(K^+\pi^-)$ &$0.1 ^{+5.3}_{-7.3}(33.2\%)$&$ -34.5^{+9.2}_{-12.2}(25.6\%)$ &$12.8 ^{+0.8}_{-1.4}(41.2\%)$&$ -3.5^{+2.2}_{-2.7}$\\
Data   & $\cdots$  &$\cdots$  &$\cdots$ &$-6\pm 9$\\
$B^0 \to\rho^-K^{*+}\to(\pi^-\pi^0)(K^0\pi^+)$  &$57.2 ^{+5.8}_{-10.1}(48.0\%)$&$ -26.5^{+5.3}_{-4.7}(25.8\%)$ &$ -30.7^{+4.6}_{-4.9}(26.2\%)$&$12.6 ^{+12.0}_{-9.5}$\\
Data   & $\cdots$  &$\cdots$  &$\cdots$ &$21\pm 15$\\
$B^0_s\to \rho^+K^{*-}\to(\pi^+\pi^0)({\bar K}^0\pi^-)$ &$ -14.2^{+3.4}_{-3.8}(89.4\%)$&$73.7 ^{+12.5}_{-15.4}(5.3\%)$ &$ 75.2^{+11.5}_{-14.1}(5.3\%)$&$-4.7 ^{+3.1}_{-2.9}$\\
$B^0_s\to\rho^0 \bar{K}^{*0}\to(\pi^+\pi^-)(K^-\pi^+)$  &$ 21.2^{+19.6}_{-16.5}(59.4\%)$&$73.2 ^{+15.4}_{-25.0}(19.1\%)$ &$ 81.8^{+11.6}_{-21.8}(21.5\%)$&$ 44.2^{+11.5}_{-13.9}$\\
$B_s^0 \to \rho^0\phi\to(\pi^+\pi^-)(K^+K^-)$  &$-2.4 ^{+9.0}_{-6.9}(82.9\%)$&$ -18.3^{+5.2}_{-4.2}(8.2\%)$ &$ -16.8^{+4.0}_{-2.9}(8.9\%)$&$-5.0 ^{+8.2}_{-5.8}$\\
$B^+ \to K^{*+}\phi\to(K^0\pi^+)(K^+K^-)$ &$ -5.7^{+11.2}_{-3.7}(54.6\%)$&$2.8 ^{+0.3}_{-2.9}(22.3\%)$ &$ -2.3^{+1.3}_{-1.0}(23.1\%)$&$ -3.3^{+7.6}_{-1.6}$\\
$B^+ \to K^{*+}{\bar K}^{*0}\to(K^0\pi^+)(K^+\pi^-)$ &$ -21.5^{+8.7}_{-10.8}(83.5\%)$&$-9.0 ^{+1.4}_{-1.3}(8.0\%)$ &$ 7.9^{+1.7}_{-1.5}(8.5\%)$&$ -19.4^{+7.6}_{-9.2}$\\
$B^0 \to K^{*+}{ K}^{*-}\to(K^0\pi^+)({\bar K}^0\pi^-)$  &$19.1 ^{+2.8}_{-1.5}(\sim 100\%)$&$ -29.5^{+17.3}_{-16.0}(\sim 0)$ &$ 10.2^{+6.7}_{-5.1}(\sim 0)$&$ 19.1^{+2.8}_{-1.5}$\\
$B_s^0 \to K^{*+}{ K}^{*-}\to(K^0\pi^+)({\bar K}^0\pi^-)$ &$33.1 ^{+7.2}_{-8.2}(32.3\%)$&$-17.4 ^{+4.5}_{-5.4}(33.8\%)$ &$ -16.6^{+4.6}_{-5.2}(33.9\%)$&$-0.9 ^{+1.6}_{-2.4}$\\
\end{tabular}}
\end{ruledtabular}
\end{table}

The direct $CP$ asymmetries from the integration over the phase space of the considered three-body $B$ meson decays are listed in Table~\ref{cpthree}.
Those modes induced only by penguin operators are expected to have vanishing $CP$ asymmetries in the LO PQCD framework and not shown in the table.
The predicted direct $CP$ asymmetry of the $B^+ \to \pi^+\rho^0$ decay is found to be negative and large, but the measured value ${\cal A}_{CP}(B^+\to \pi^+\rho^0)=(-0.4\pm 1.7\pm0.9)\%$ is consistent with zero~\cite{LHCb:2022tuk}.
This mismatch represents a puzzle that needs to be resolved.
The CPV originating from the interference between the $S$- and $P$-wave $\pi^+\pi^-$ pairs in the $B^+\to \pi^+\rho^0$ decay has been observed by the LHCb~\cite{LHCb:2019jta}.
The interference causes significant local asymmetries, which, however, cancel in the integration over the phase space of the decay.
It has been postulated~\cite{Cheng:2020hyj,Cheng:2022ysn} that the nearly diminishing CPV is due to the destruction between the power corrections from the penguin annihilation and hard spectator interactions.

The LHCb Collaboration reported the first observation of the direct $CP$ asymmetry ${\cal A}_{CP}(B^+\to K^+\rho^0)=(15.0\pm 1.9\pm1.1)\%$~\cite{LHCb:2022tuk} recently based on the measurements of the three-body decay $B^\pm \to K^\pm \pi^+\pi^-$.
Our prediction ${\cal A}_{CP}(B^+\to K^+\rho^0)=(62.4^{+11.6}_{-13.5})\%$, similar to the one from the formalism for two-body decays~\cite{Chai:2022ptk}, is higher than the data.
The authors of Ref.~\cite{Cheng:2014rka} noticed that NLO corrections to the scalar pion form factor are small in magnitude but generate a large strong phase, which can affect dramatically direct $CP$ asymmetries.
The discrepancy between our prediction and the data for ${\cal A}_{CP}(B^+\to K^+\rho^0)$ may thus be removed, when complete NLO corrections to multi-body $B$ decays, in particular those to nonfactorizable spectator and annihilation diagrams, are available.

$CP$ asymmetries of the $B \to V_1V_2$ modes have not been observed at a level higher than $5\sigma$ so far, most of which exhibit no apparent deviation from zero.
There are three possible spin configurations corresponding to the polarizations of the final-state vector mesons $V_1$ and $V_2$, which are allowed by the angular momentum conservation, while there is only one helicity amplitude in $B$ meson decays into two pseudoscalars.
We demonstrate below that $CP$ asymmetries in individual helicity final states of some $B \to V_1V_2$ decays can be as large as those in the $B \to P_1P_2$ ones, greater than $10\%$, but the cancellation between different polarization components results in small net CPV.

For the four-body decays $B\to V_1V_2\to (P_1P_2)(P_3P_4)$, one can disentangle the helicities of the $V_1(\to P_1P_2)V_2(\to P_3P_4)$ final states via an angular analysis.
We collect in Table~\ref{cpfour} the direct $CP$ asymmetries in the three helicity states ${\cal A}^{0,||,\bot}_{CP}$ together with ${\cal A}^{\rm dir}_{ CP}$ from the summation over all helicity states.
The total direct $CP$ asymmetry can be well approximated by the
weighted sum of the three asymmetries,
\begin{eqnarray}\label{appr}
{\cal A}^{\rm dir}_{ CP}\approx f_0{\cal A}^{0}_{CP}+f_{||}{\cal A}^{||}_{CP}+f_{\bot}{\cal A}^{\bot}_{CP},
\end{eqnarray}
with the coefficients $f_i$,
$i=0,||,\bot$, being the corresponding polarization fractions.
The predicted ${\cal A}^{||}_{ CP}$ for the $B^0 \to \rho^0K^{*0}\to(\pi^+\pi^-)(K^+\pi^-)$ channel can reach $(-34.5^{+9.2}_{-12.2})\%$,
but the cancellation among the three helicity components in Eq.~(\ref{appr}) decreases the magnitude of ${\cal A}^{\rm dir}_{ CP}$, rendering it compatible with the data.
We mention that similar cancellation between various partial waves also appears in the analysis of the CPV in bottom baryon decays~\cite{Yu:2024cjd}.
The tiny asymmetry ${\cal A}^{0}_{CP}$ in the longitudinal component of the $B_s^0 \to \rho^0\phi\to(\pi^+\pi^-)(K^+K^-)$ decay governs the direct CPV owing to the weight of nearly unity, $f_0\sim 80\%$, giving rise to small ${\cal A}^{\rm dir}_{ CP}$.
The direct $CP$ asymmetries of the other decays can be understood in the same manner.


\section{CONCLUSION}

We have improved the PQCD formalism of multi-body charmless hadronic $B$ meson decays by resolving the possible inconsistency in the parametrization for $P$-wave resonances in two-meson DAs.
The determination of the Gegenbauer moments in two-meson DAs was then updated by performing the global fit to measured branching ratios and polarization fractions for the three-body $B\to VP_3\to P_1P_2P_3$ and four-body $B\to V_1V_2 \to (P_1P_2)(P_3P_4)$ decays in the LO PQCD framework.
The data from the three additional four-body decays $B \to \rho K^*\to(\pi\pi)(K\pi)$, $B \to K^* K^*\to(K\pi)(K\pi)$  and $B \to \phi K^*\to(KK)(K\pi)$ were included to reduce theoretical uncertainties.
The moments $a^{\perp}_{1K^*}$ and $a^{\perp}_{2K^*}$ in the transverse $K\pi$ DA were extracted from the global analysis for the first time.
It has been shown that the resultant Gegenbauer expansion of two-meson DAs becomes more convergent.

With the updated two-meson DAs, the branching ratios, polarization fractions and the direct $CP$ asymmetries of the considered multi-body $B$ meson decays were reexamined.
It has been demonstrated that most of the data involved in the fit are well reproduced; namely, the fit quality is satisfactory.
The precision of the two-meson DAs does play a crucial role in accounting for the data, especially for the unexpected low longitudinal polarization fraction of the $B_s^0\to K^{*0} {\bar K}^{*0}$ decay.
We have predicted the observable $L_{K^*K^*}$ in the PQCD approach, whose confrontation with measurements can probe the presence of NP contributions.
It was also found that although the direct $CP$ asymmetry in each polarization component of some four-body $B$ meson decays might be sizable, the strong cancellation among polarization states leads to diminishing net CPV. Therefore, it is worthwhile to search for CPV in individual polarization states experimentally.


The present work is still based on the LO PQCD factorization formulas.
The precision of the two-meson DAs can be further enhanced systematically, when higher-order and/or higher-power corrections to multi-body hadronic $B$-meson decays are taken into account.
More precise data, in particular those for $CP$ asymmetries, are also necessary.
If a high-precision global investigation discloses notable tensions between theoretical results and experimental data, it may hint that NP effects are inevitable.

\begin{acknowledgments}
We thank Fu-Sheng Yu, Rui-Lin Zhu and Jia-Jie Han for valuable discussions.
This work is supported by the National Natural Science Foundation of China under
Grant Nos.~12005103, 12075086, and by NSTC of R.O.C. under Grant No. NSTC-113-2112-M-001-024-MY3.
DCY is supported by the Natural Science Foundation of Jiangsu Province under Grant No.~BK20200980.
ZR is supported in part by the Natural Science Foundation of Hebei Province under Grant Nos.~A2019209449 and A2021209002.
\end{acknowledgments}



\begin{thebibliography}{199}

\bibitem{Dalitz:1953cp}
R.~H.~Dalitz,
Phil. Mag. Ser. 7 \textbf{44} (1953), 1068-1080.
doi:10.1080/14786441008520365

\bibitem{Dalitz:1954cq}
R.~H.~Dalitz,
Phys. Rev. \textbf{94} (1954), 1046-1051
doi:10.1103/PhysRev.94.1046















\bibitem{LHCb:2013ptu}
R.~Aaij \textit{et al.} [LHCb],
Phys. Rev. Lett. \textbf{111} (2013), 101801
doi:10.1103/PhysRevLett.111.101801
[arXiv:1306.1246 [hep-ex]].


\bibitem{LHCb:2013lcl}
R.~Aaij \textit{et al.} [LHCb],
Phys. Rev. Lett. \textbf{112} (2014) no.1, 011801
doi:10.1103/PhysRevLett.112.011801
[arXiv:1310.4740 [hep-ex]].

\bibitem{LHCb:2022fpg}
R.~Aaij \textit{et al.} [LHCb],
Phys. Rev. D \textbf{108} (2023) no.1, 012008
doi:10.1103/PhysRevD.108.012008
[arXiv:2206.07622 [hep-ex]].


\bibitem{LHCb:2019jta}
R.~Aaij \textit{et al.} [LHCb],
Phys. Rev. Lett. \textbf{124} (2020) no.3, 031801
doi:10.1103/PhysRevLett.124.031801
[arXiv:1909.05211 [hep-ex]].



\bibitem{Bediaga:2013ela}
I.~Bediaga, T.~Frederico and O.~Louren\c{c}o,
Phys. Rev. D \textbf{89} (2014) no.9, 094013
doi:10.1103/PhysRevD.89.094013
[arXiv:1307.8164 [hep-ph]].

\bibitem{Bediaga:2015mia}
I.~Bediaga and P.~C.~Magalh\~aes,
[arXiv:1512.09284 [hep-ph]].

\bibitem{Kang:2013jaa}
X.~W.~Kang, B.~Kubis, C.~Hanhart and U.~G.~Mei\ss{}ner,
Phys. Rev. D \textbf{89} (2014), 053015
doi:10.1103/PhysRevD.89.053015
[arXiv:1312.1193 [hep-ph]].


\bibitem{Chen:2002th}
C.~H.~Chen and H.~n.~Li,
Phys. Lett. B \textbf{561} (2003), 258-265
doi:10.1016/S0370-2693(03)00486-6
[arXiv:hep-ph/0209043 [hep-ph]].


\bibitem{El-Bennich:2009gqk}
B.~El-Bennich, A.~Furman, R.~Kaminski, L.~Lesniak, B.~Loiseau and B.~Moussallam,
Phys. Rev. D \textbf{79} (2009), 094005
[erratum: Phys. Rev. D \textbf{83} (2011), 039903]
doi:10.1103/PhysRevD.83.039903
[arXiv:0902.3645 [hep-ph]].

\bibitem{Virto:2016fbw}
J.~Virto,
PoS \textbf{FPCP2016} (2017), 007
doi:10.22323/1.280.0007
[arXiv:1609.07430 [hep-ph]].


\bibitem{Krankl:2015fha}
S.~Kr\"ankl, T.~Mannel and J.~Virto,
Nucl. Phys. B \textbf{899} (2015), 247-264
doi:10.1016/j.nuclphysb.2015.08.004
[arXiv:1505.04111 [hep-ph]].

\bibitem{G}
A.~G.~Grozin,
Sov. J. Nucl. Phys. \textbf{38} (1983), 289-292

\bibitem{G1}
A.~G.~Grozin,
Theor. Math. Phys. \textbf{69} (1986), 1109-1121
doi:10.1007/BF01037870

\bibitem{DM}
D.~M\"uller, D.~Robaschik, B.~Geyer, F.~M.~Dittes and J.~Ho\v{r}ej\v{s}i,
Fortsch. Phys. \textbf{42} (1994), 101-141
doi:10.1002/prop.2190420202
[arXiv:hep-ph/9812448 [hep-ph]].

\bibitem{Diehl:1998dk}
M.~Diehl, T.~Gousset, B.~Pire and O.~Teryaev,
Phys. Rev. Lett. \textbf{81} (1998), 1782-1785
doi:10.1103/PhysRevLett.81.1782
[arXiv:hep-ph/9805380 [hep-ph]].

\bibitem{Diehl:1998dk1}
M.~Diehl, T.~Gousset and B.~Pire,
Phys. Rev. D \textbf{62} (2000), 073014
doi:10.1103/PhysRevD.62.073014
[arXiv:hep-ph/0003233 [hep-ph]].

\bibitem{Diehl:1998dk2}
B.~Pire and L.~Szymanowski,
Phys. Lett. B \textbf{556} (2003), 129-134
doi:10.1016/S0370-2693(03)00134-5
[arXiv:hep-ph/0212296 [hep-ph]].



\bibitem{MP}
M.~V.~Polyakov,
Nucl. Phys. B \textbf{555} (1999), 231
doi:10.1016/S0550-3213(99)00314-4
[arXiv:hep-ph/9809483 [hep-ph]].





\bibitem{Li:2021cnd}
Y.~Li, D.~C.~Yan, J.~Hua, Z.~Rui and H.~n.~Li,
Phys. Rev. D \textbf{104} (2021) no.9, 096014
doi:10.1103/PhysRevD.104.096014
[arXiv:2105.03899 [hep-ph]].





\bibitem{Hua:2020usv}
J.~Hua, H.~n.~Li, C.~D.~Lu, W.~Wang and Z.~P.~Xing,
Phys. Rev. D \textbf{104} (2021) no.1, 016025
doi:10.1103/PhysRevD.104.016025
[arXiv:2012.15074 [hep-ph]].


\bibitem{Rui:2021kbn}
Z.~Rui, Y.~Li and H.~n.~Li,
JHEP \textbf{05}, 082 (2021)
doi:10.1007/JHEP05(2021)082
[arXiv:2103.00642 [hep-ph]].

\bibitem{Li:2021qiw}
Y.~Li, D.~C.~Yan, Z.~Rui and Z.~J.~Xiao,
Eur. Phys. J. C \textbf{81}, no.9, 806 (2021)
doi:10.1140/epjc/s10052-021-09608-5
[arXiv:2107.10684 [hep-ph]].

\bibitem{Zhang:2021nlw}
C.~Q.~Zhang, J.~M.~Li, M.~K.~Jia, Y.~Li and Z.~Rui,
Phys. Rev. D \textbf{105}, no.5, 053002 (2022)
doi:10.1103/PhysRevD.105.053002
[arXiv:2112.10939 [hep-ph]].

\bibitem{Yan:2022kck}
D.~C.~Yan, Z.~Rui, Z.~J.~Xiao and Y.~Li,
Phys. Rev. D \textbf{105}, no.9, 093001 (2022)
doi:10.1103/PhysRevD.105.093001
[arXiv:2204.01092 [hep-ph]].

\bibitem{Zhang:2022pfn}
Z.~Q.~Zhang, Y.~C.~Zhao, Z.~L.~Guan, Z.~J.~Sun, Z.~Y.~Zhang and K.~Y.~He,
Chin. Phys. C \textbf{46}, no.12, 123105 (2022)
doi:10.1088/1674-1137/ac89d1
[arXiv:2207.02043 [hep-ph]].

\bibitem{Zou:2022xrr}
Z.~T.~Zou, W.~S.~Fang, X.~Liu and Y.~Li,
Eur. Phys. J. C \textbf{82}, no.11, 1076 (2022)
doi:10.1140/epjc/s10052-022-11060-y
[arXiv:2210.08522 [hep-ph]].

\bibitem{Yan:2023yvx}
D.~C.~Yan, Z.~Rui, Y.~Yan and Y.~Li,
Eur. Phys. J. C \textbf{83}, no.10, 974 (2023)
doi:10.1140/epjc/s10052-023-12152-z
[arXiv:2308.12543 [hep-ph]].

\bibitem{Yan:2024ymv}
D.~C.~Yan, Y.~Yan and Z.~Rui,
Eur. Phys. J. C \textbf{84}, no.7, 754 (2024)
doi:10.1140/epjc/s10052-024-13087-9
[arXiv:2404.19198 [hep-ph]].

\bibitem{Dedonder:2010fg}
J.~P.~Dedonder, A.~Furman, R.~Kaminski, L.~Lesniak and B.~Loiseau,
Acta Phys. Polon. B \textbf{42} (2011), 2013
doi:10.5506/APhysPolB.42.2013
[arXiv:1011.0960 [hep-ph]].

\bibitem{Alguero:2020xca}
M.~Alguer\'o, A.~Crivellin, S.~Descotes-Genon, J.~Matias and M.~Novoa-Brunet,
JHEP \textbf{04}, 066 (2021)
doi:10.1007/JHEP04(2021)066
[arXiv:2011.07867 [hep-ph]].











\bibitem{LHCb:2018oeg}
R.~Aaij \textit{et al.} [LHCb],
Eur. Phys. J. C \textbf{78} (2018) no.12, 1019
doi:10.1140/epjc/s10052-018-6447-z
[arXiv:1809.07416 [hep-ex]].

\bibitem{Gounaris:1968mw}
G.~J.~Gounaris and J.~J.~Sakurai,
Phys. Rev. Lett. \textbf{21} (1968), 244-247
doi:10.1103/PhysRevLett.21.244

\bibitem{Watson:1952ji}
K.~M.~Watson,
Phys. Rev. \textbf{88} (1952), 1163-1171
doi:10.1103/PhysRev.88.1163

















\bibitem{pdg2024}
S.~Navas \textit{et al.} [Particle Data Group],
Phys. Rev. D \textbf{110} (2024) no.3, 030001
doi:10.1103/PhysRevD.110.030001


\bibitem{prd63-054008}
Y.~Y.~Keum, H.~N.~Li and A.~I.~Sanda,
Phys. Rev. D \textbf{63} (2001), 054008
doi:10.1103/PhysRevD.63.054008
[arXiv:hep-ph/0004173 [hep-ph]].

\bibitem{prd65-014007}
T.~Kurimoto, H.~n.~Li and A.~I.~Sanda,
Phys. Rev. D \textbf{65} (2002), 014007
doi:10.1103/PhysRevD.65.014007
[arXiv:hep-ph/0105003 [hep-ph]].

\bibitem{epjc28-515}
C.~D.~Lu and M.~Z.~Yang,
Eur. Phys. J. C \textbf{28} (2003), 515-523
doi:10.1140/epjc/s2003-01199-y
[arXiv:hep-ph/0212373 [hep-ph]].

\bibitem{ppnp51-85}
H.~n.~Li,
Prog. Part. Nucl. Phys. \textbf{51} (2003), 85-171
doi:10.1016/S0146-6410(03)90013-5
[arXiv:hep-ph/0303116 [hep-ph]].

\bibitem{prd85-094003}
Z.~J.~Xiao, W.~F.~Wang and Y.~y.~Fan,
Phys. Rev. D \textbf{85} (2012), 094003
doi:10.1103/PhysRevD.85.094003
[arXiv:1111.6264 [hep-ph]].

\bibitem{Li:2012md}
H.~N.~Li, Y.~L.~Shen and Y.~M.~Wang,
JHEP \textbf{02} (2013), 008
doi:10.1007/JHEP02(2013)008
[arXiv:1210.2978 [hep-ph]].








\bibitem{prd63-074009}
C.~D.~Lu, K.~Ukai and M.~Z.~Yang,
Phys. Rev. D \textbf{63} (2001), 074009
doi:10.1103/PhysRevD.63.074009
[arXiv:hep-ph/0004213 [hep-ph]].

\bibitem{plb504-6}
Y.~Y.~Keum, H.~n.~Li and A.~I.~Sanda,
Phys. Lett. B \textbf{504} (2001), 6-14
doi:10.1016/S0370-2693(01)00247-7
[arXiv:hep-ph/0004004 [hep-ph]].


\bibitem{Grozin:1996pq}
A.~G.~Grozin and M.~Neubert,
Phys. Rev. D \textbf{55} (1997), 272-290
doi:10.1103/PhysRevD.55.272
[arXiv:hep-ph/9607366 [hep-ph]].




\bibitem{prd103-056006}
Y.~Yang, L.~Lang, X.~Zhao, J.~Huang and J.~Sun,
Phys. Rev. D \textbf{103} (2021) no.5, 056006
doi:10.1103/PhysRevD.103.056006
[arXiv:2012.10581 [hep-ph]].

\bibitem{jhep05-022}
V.~M.~Braun, Y.~Ji and A.~N.~Manashov,
JHEP \textbf{05} (2017), 022
doi:10.1007/JHEP05(2017)022
[arXiv:1703.02446 [hep-ph]].


\bibitem{prd76-074018}
A.~Ali, G.~Kramer, Y.~Li, C.~D.~Lu, Y.~L.~Shen, W.~Wang and Y.~M.~Wang,
Phys. Rev. D \textbf{76} (2007), 074018
doi:10.1103/PhysRevD.76.074018
[arXiv:hep-ph/0703162 [hep-ph]].

\bibitem{plb763-29}
W.~F.~Wang and H.~n.~Li,
Phys. Lett. B \textbf{763} (2016), 29-39
doi:10.1016/j.physletb.2016.10.026
[arXiv:1609.04614 [hep-ph]].

\bibitem{Rui:2018hls}
Z.~Rui, Y.~Li and H.~N.~Li,
Phys. Rev. D \textbf{98} (2018) no.11, 113003
doi:10.1103/PhysRevD.98.113003
[arXiv:1809.04754 [hep-ph]].




\bibitem{LHCb:2018hsm}
R.~Aaij \textit{et al.} [LHCb],
JHEP \textbf{05}, 026 (2019)
doi:10.1007/JHEP05(2019)026
[arXiv:1812.07008 [hep-ex]].



\bibitem{BW-model}
G.~Breit and E.~Wigner,
Phys. Rev. \textbf{49} (1936), 519-531
doi:10.1103/PhysRev.49.519





\bibitem{prd95056008}
Y.~Li, A.~J.~Ma, W.~F.~Wang and Z.~J.~Xiao,
Phys. Rev. D \textbf{95} (2017) no.5, 056008
doi:10.1103/PhysRevD.95.056008
[arXiv:1612.05934 [hep-ph]].


\bibitem{prd86-032013}
J.~P.~Lees \textit{et al.} [BaBar],
Phys. Rev. D \textbf{86} (2012), 032013
doi:10.1103/PhysRevD.86.032013
[arXiv:1205.2228 [hep-ex]].


\bibitem{Ali:1998eb}
A.~Ali, G.~Kramer and C.~D.~Lu,
Phys. Rev. D \textbf{58} (1998), 094009
doi:10.1103/PhysRevD.58.094009
[arXiv:hep-ph/9804363 [hep-ph]].




\bibitem{Cheng:2022ysn}
H.~Y.~Cheng,
Phys. Rev. D \textbf{106} (2022) no.11, 113004
doi:10.1103/PhysRevD.106.113004
[arXiv:2211.03965 [hep-ph]].








\bibitem{Peter:2020}
P.~Lepage and C.~Gohlke,
gplepage/lsqfit: lsqfit version 11.7,  Zenodo. http://doi.org/10.5281/zenodo.4037174.


\bibitem{Li:2006jv}
H.~n.~Li and S.~Mishima,
Phys. Rev. D \textbf{74} (2006), 094020
doi:10.1103/PhysRevD.74.094020
[arXiv:hep-ph/0608277 [hep-ph]].

\bibitem{Rui:2011dr}
Z.~Rui, X.~Gao and C.~D.~Lu,
Eur. Phys. J. C \textbf{72} (2012), 1923
doi:10.1140/epjc/s10052-012-1923-3
[arXiv:1111.0181 [hep-ph]].

\bibitem{ball98}
P.~Ball, V.~M.~Braun, Y.~Koike and K.~Tanaka,
Nucl. Phys. B \textbf{529} (1998), 323-382
doi:10.1016/S0550-3213(98)00356-3
[arXiv:hep-ph/9802299 [hep-ph]].

\bibitem{Liu:2015sra}
X.~Liu, H.~n.~Li and Z.~J.~Xiao,
Phys. Rev. D \textbf{91}, no.11, 114019 (2015)
doi:10.1103/PhysRevD.91.114019
[arXiv:1502.04162 [hep-ph]].




\bibitem{Wang:2015vgv}
Y.~M.~Wang and Y.~L.~Shen,
Nucl. Phys. B \textbf{898} (2015), 563-604
doi:10.1016/j.nuclphysb.2015.07.016
[arXiv:1506.00667 [hep-ph]].


\bibitem{Wang:2016qii}
Y.~M.~Wang,
JHEP \textbf{09} (2016), 159
doi:10.1007/JHEP09(2016)159
[arXiv:1606.03080 [hep-ph]].

\bibitem{Wang:2018wfj}
Y.~M.~Wang and Y.~L.~Shen,
JHEP \textbf{05} (2018), 184
doi:10.1007/JHEP05(2018)184
[arXiv:1803.06667 [hep-ph]].


\bibitem{Wang:2019msf}
W.~Wang, Y.~M.~Wang, J.~Xu and S.~Zhao,
Phys. Rev. D \textbf{102} (2020) no.1, 011502
doi:10.1103/PhysRevD.102.011502
[arXiv:1908.09933 [hep-ph]].


\bibitem{Han:2024min}
X.~Y.~Han, J.~Hua, X.~Ji, C.~D.~L\"u, W.~Wang, J.~Xu, Q.~A.~Zhang and S.~Zhao,
[arXiv:2403.17492 [hep-ph]].

\bibitem{LatticeParton:2024zko}
X.~Y.~Han \textit{et al.} [Lattice Parton],
Phys. Rev. D \textbf{111} (2025) no.3, 034503
doi:10.1103/PhysRevD.111.034503
[arXiv:2410.18654 [hep-lat]].


\bibitem{Yan:2017nlj}
D.~C.~Yan, P.~Yang, X.~Liu and Z.~J.~Xiao,
Nucl. Phys. B \textbf{931}, 79-104 (2018)
doi:10.1016/j.nuclphysb.2018.04.007
[arXiv:1707.06043 [hep-ph]].

\bibitem{Gubernari:2018wyi}
N.~Gubernari, A.~Kokulu and D.~van Dyk,
JHEP \textbf{01}, 150 (2019)
doi:10.1007/JHEP01(2019)150
[arXiv:1811.00983 [hep-ph]].

\bibitem{Khodjamirian:2006st}
A.~Khodjamirian, T.~Mannel and N.~Offen,
Phys. Rev. D \textbf{75}, 054013 (2007)
doi:10.1103/PhysRevD.75.054013
[arXiv:hep-ph/0611193 [hep-ph]].


\bibitem{Bharucha:2015bzk}
A.~Bharucha, D.~M.~Straub and R.~Zwicky,
JHEP \textbf{08}, 098 (2016)
doi:10.1007/JHEP08(2016)098
[arXiv:1503.05534 [hep-ph]].

\bibitem{Cheng:2017smj}
S.~Cheng, A.~Khodjamirian and J.~Virto,
JHEP \textbf{05}, 157 (2017)
doi:10.1007/JHEP05(2017)157
[arXiv:1701.01633 [hep-ph]].

\bibitem{Ball:2004rg}
P.~Ball and R.~Zwicky,
Phys. Rev. D \textbf{71}, 014029 (2005)
doi:10.1103/PhysRevD.71.014029
[arXiv:hep-ph/0412079 [hep-ph]].

\bibitem{Khodjamirian:2010vf}
A.~Khodjamirian, T.~Mannel, A.~A.~Pivovarov and Y.~M.~Wang,
JHEP \textbf{09}, 089 (2010)
doi:10.1007/JHEP09(2010)089
[arXiv:1006.4945 [hep-ph]].

\bibitem{Descotes-Genon:2019bud}
S.~Descotes-Genon, A.~Khodjamirian and J.~Virto,
JHEP \textbf{12}, 083 (2019)
doi:10.1007/JHEP12(2019)083
[arXiv:1908.02267 [hep-ph]].




\bibitem{Cheng:2009mu}
H.~Y.~Cheng and C.~K.~Chua,
Phys. Rev. D \textbf{80} (2009), 114026
doi:10.1103/PhysRevD.80.114026
[arXiv:0910.5237 [hep-ph]].

\bibitem{Wang:2017rmh}
C.~Wang, S.~H.~Zhou, Y.~Li and C.~D.~Lu,
Phys. Rev. D \textbf{96} (2017) no.7, 073004
doi:10.1103/PhysRevD.96.073004
[arXiv:1708.04861 [hep-ph]].


\bibitem{Wang:2017hxe}
C.~Wang, Q.~A.~Zhang, Y.~Li and C.~D.~Lu,
Eur. Phys. J. C \textbf{77} (2017) no.5, 333
doi:10.1140/epjc/s10052-017-4889-3
[arXiv:1701.01300 [hep-ph]].





\bibitem{prd91-054033}
Z.~T.~Zou, A.~Ali, C.~D.~Lu, X.~Liu and Y.~Li,
Phys. Rev. D \textbf{91} (2015), 054033
doi:10.1103/PhysRevD.91.054033
[arXiv:1501.00784 [hep-ph]].



\bibitem{plb622-63}
H.~n.~Li,
Phys. Lett. B \textbf{622} (2005), 63-68
doi:10.1016/j.physletb.2005.06.077
[arXiv:hep-ph/0411305 [hep-ph]].

\bibitem{prl107-261802}
T.~Aaltonen \textit{et al.} [CDF],
Phys. Rev. Lett. \textbf{107} (2011), 261802
doi:10.1103/PhysRevLett.107.261802
[arXiv:1107.4999 [hep-ex]].

\bibitem{plb713-369}
R.~Aaij \textit{et al.} [LHCb],
Phys. Lett. B \textbf{713} (2012), 369-377
doi:10.1016/j.physletb.2012.06.012
[arXiv:1204.2813 [hep-ex]].

\bibitem{prl91-201801}
K.~F.~Chen \textit{et al.} [Belle],
Phys. Rev. Lett. \textbf{91} (2003), 201801
doi:10.1103/PhysRevLett.91.201801
[arXiv:hep-ex/0307014 [hep-ex]].

\bibitem{prd78-092008}
B.~Aubert \textit{et al.} [BaBar],
Phys. Rev. D \textbf{78} (2008), 092008
doi:10.1103/PhysRevD.78.092008
[arXiv:0808.3586 [hep-ex]].

\bibitem{prd85-072005}
J.~P.~Lees \textit{et al.} [BaBar],
Phys. Rev. D \textbf{85} (2012), 072005
doi:10.1103/PhysRevD.85.072005
[arXiv:1112.3896 [hep-ex]].

\bibitem{npb774-64}
M.~Beneke, J.~Rohrer and D.~Yang,
Nucl. Phys. B \textbf{774} (2007), 64-101
doi:10.1016/j.nuclphysb.2007.03.020
[arXiv:hep-ph/0612290 [hep-ph]].

\bibitem{prd70-054015}
C.~W.~Bauer, D.~Pirjol, I.~Z.~Rothstein and I.~W.~Stewart,
Phys. Rev. D \textbf{70} (2004), 054015
doi:10.1103/PhysRevD.70.054015
[arXiv:hep-ph/0401188 [hep-ph]].

\bibitem{prd71-054025}
H.~n.~Li and S.~Mishima,
Phys. Rev. D \textbf{71} (2005), 054025
doi:10.1103/PhysRevD.71.054025
[arXiv:hep-ph/0411146 [hep-ph]].

\bibitem{Cheng:2008gxa}
H.~Y.~Cheng and K.~C.~Yang,
Phys. Rev. D \textbf{78} (2008), 094001
[erratum: Phys. Rev. D \textbf{79} (2009), 039903]
doi:10.1103/PhysRevD.79.039903
[arXiv:0805.0329 [hep-ph]].

\bibitem{Grossman:2003qi}
Y.~Grossman,
Int. J. Mod. Phys. A \textbf{19} (2004), 907-917
doi:10.1142/S0217751X04018865
[arXiv:hep-ph/0310229 [hep-ph]].

\bibitem{Das:2004hq}
P.~K.~Das and K.~C.~Yang,
Phys. Rev. D \textbf{71} (2005), 094002
doi:10.1103/PhysRevD.71.094002
[arXiv:hep-ph/0412313 [hep-ph]].

\bibitem{Chen:2005mka}
C.~H.~Chen and C.~Q.~Geng,
Phys. Rev. D \textbf{71} (2005), 115004
doi:10.1103/PhysRevD.71.115004
[arXiv:hep-ph/0504145 [hep-ph]].

\bibitem{Yang:2004pm}
Y.~D.~Yang, R.~M.~Wang and G.~R.~Lu,
Phys. Rev. D \textbf{72} (2005), 015009
doi:10.1103/PhysRevD.72.015009
[arXiv:hep-ph/0411211 [hep-ph]].

\bibitem{Kagan:2004uw}
A.~L.~Kagan,
Phys. Lett. B \textbf{601} (2004), 151-163
doi:10.1016/j.physletb.2004.09.030
[arXiv:hep-ph/0405134 [hep-ph]].

\bibitem{Beneke:2005we}
M.~Beneke, J.~Rohrer and D.~Yang,
Phys. Rev. Lett. \textbf{96} (2006), 141801
doi:10.1103/PhysRevLett.96.141801
[arXiv:hep-ph/0512258 [hep-ph]].

\bibitem{Datta:2007qb}
A.~Datta, A.~V.~Gritsan, D.~London, M.~Nagashima and A.~Szynkman,
Phys. Rev. D \textbf{76} (2007), 034015
doi:10.1103/PhysRevD.76.034015
[arXiv:0705.3915 [hep-ph]].

\bibitem{Colangelo:2004rd}
P.~Colangelo, F.~De Fazio and T.~N.~Pham,
Phys. Lett. B \textbf{597} (2004), 291-298
doi:10.1016/j.physletb.2004.07.024
[arXiv:hep-ph/0406162 [hep-ph]].

\bibitem{Ladisa:2004bp}
M.~Ladisa, V.~Laporta, G.~Nardulli and P.~Santorelli,
Phys. Rev. D \textbf{70} (2004), 114025
doi:10.1103/PhysRevD.70.114025
[arXiv:hep-ph/0409286 [hep-ph]].

\bibitem{Cheng:2004ru}
H.~Y.~Cheng, C.~K.~Chua and A.~Soni,
Phys. Rev. D \textbf{71} (2005), 014030
doi:10.1103/PhysRevD.71.014030
[arXiv:hep-ph/0409317 [hep-ph]].
















\bibitem{Alvarez:2004ci}
E.~Alvarez, L.~N.~Epele, D.~Gomez Dumm and A.~Szynkman,
Phys. Rev. D \textbf{70} (2004), 115014
doi:10.1103/PhysRevD.70.115014
[arXiv:hep-ph/0410096 [hep-ph]].

\bibitem{Yang:2005tv}
K.~C.~Yang,
Phys. Rev. D \textbf{72} (2005), 034009
[erratum: Phys. Rev. D \textbf{72} (2005), 059901]
doi:10.1103/PhysRevD.72.034009
[arXiv:hep-ph/0506040 [hep-ph]].

\bibitem{Baek:2005jk}
S.~Baek, A.~Datta, P.~Hamel, O.~F.~Hernandez and D.~London,
Phys. Rev. D \textbf{72} (2005), 094008
doi:10.1103/PhysRevD.72.094008
[arXiv:hep-ph/0508149 [hep-ph]].

\bibitem{Huang:2005qb}
C.~S.~Huang, P.~Ko, X.~H.~Wu and Y.~D.~Yang,
Phys. Rev. D \textbf{73} (2006), 034026
doi:10.1103/PhysRevD.73.034026
[arXiv:hep-ph/0511129 [hep-ph]].

\bibitem{Chen:2006vs}
C.~H.~Chen and H.~Hatanaka,
Phys. Rev. D \textbf{73} (2006), 075003
doi:10.1103/PhysRevD.73.075003
[arXiv:hep-ph/0602140 [hep-ph]].

\bibitem{Faessler:2007br}
A.~Faessler, T.~Gutsche, J.~C.~Helo, S.~Kovalenko and V.~E.~Lyubovitskij,
Phys. Rev. D \textbf{75} (2007), 074029
doi:10.1103/PhysRevD.75.074029
[arXiv:hep-ph/0702020 [hep-ph]].

\bibitem{Chen:2005cx}
C.~H.~Chen, C.~Q.~Geng, Y.~K.~Hsiao and Z.~T.~Wei,
Phys. Rev. D \textbf{72} (2005), 054011
doi:10.1103/PhysRevD.72.054011
[arXiv:hep-ph/0507012 [hep-ph]].

\bibitem{Chen:2007qj}
C.~H.~Chen and C.~Q.~Geng,
Phys. Rev. D \textbf{75} (2007), 054010
doi:10.1103/PhysRevD.75.054010
[arXiv:hep-ph/0701023 [hep-ph]].

\bibitem{Cheng:2010yd}
H.~Y.~Cheng and K.~C.~Yang,
Phys. Rev. D \textbf{83} (2011), 034001
doi:10.1103/PhysRevD.83.034001
[arXiv:1010.3309 [hep-ph]].

\bibitem{Bobeth:2014rra}
C.~Bobeth, M.~Gorbahn and S.~Vickers,
Eur. Phys. J. C \textbf{75} (2015) no.7, 340
doi:10.1140/epjc/s10052-015-3535-1
[arXiv:1409.3252 [hep-ph]].


\bibitem{Li:2006cva}
H.~n.~Li and S.~Mishima,
Phys. Rev. D \textbf{73}, 114014 (2006)
doi:10.1103/PhysRevD.73.114014
[arXiv:hep-ph/0602214 [hep-ph]].





\bibitem{Li:2022mtc}
Y.~Li, G.~H.~Zhao, Y.~J.~Sun and Z.~T.~Zou,
Phys. Rev. D \textbf{106}, no.9, 093009 (2022)
doi:10.1103/PhysRevD.106.093009
[arXiv:2209.13389 [hep-ph]].

\bibitem{LHCb:2022tuk}
R.~Aaij \textit{et al.} [LHCb],
Phys. Rev. D \textbf{108}, no.1, 012013 (2023)
doi:10.1103/PhysRevD.108.012013
[arXiv:2206.02038 [hep-ex]].

\bibitem{Cheng:2020hyj}
H.~Y.~Cheng,
[arXiv:2005.06080 [hep-ph]].



\bibitem{Chai:2022ptk}
J.~Chai, S.~Cheng, Y.~h.~Ju, D.~C.~Yan, C.~D.~L\"u and Z.~J.~Xiao,
Chin. Phys. C \textbf{46}, no.12, 123103 (2022)
doi:10.1088/1674-1137/ac88bd
[arXiv:2207.04190 [hep-ph]].



\bibitem{Cheng:2014rka}
S.~Cheng, Z.~J.~Xiao and Y.~L.~Zhang,
Nucl. Phys. B \textbf{896}, 255-280 (2015)
doi:10.1016/j.nuclphysb.2015.04.021
[arXiv:1409.5947 [hep-ph]].

\bibitem{Yu:2024cjd}
J.~X.~Yu, J.~J.~Han, Y.~Li, H.~n.~Li, Z.~J.~Xiao and F.~S.~Yu,
[arXiv:2409.02821 [hep-ph]].






\end{thebibliography}
\end{document}